\documentclass[twocolumn]{aastex63}

\usepackage{amsmath}
\usepackage{amssymb}
\usepackage{rotating}

\usepackage[T1]{fontenc}
\usepackage[utf8]{inputenc}

\newcommand{\ra}[4]{${#1}^{\rm h}{#2}^{\rm m}{#3}\fs{#4}$}
\newcommand{\dec}[4]{${#1}\arcdeg{#2}\arcmin{#3}\farcs{#4}$}

\newcommand\tE{t_{\rm E}}

\newcommand\piEN{\pi_{\textrm{E},N}}
\newcommand\piEE{\pi_{\textrm{E},E}}

\newcommand\pirel{\pi_{\rm rel}}
\newcommand\thetaE{\theta_{\rm E}}
\newcommand\piE{\pi_{\rm E}}
\newcommand\piEvec{\boldsymbol{\pi}_{\rm E}}

\received{}
\revised{}
\accepted{}

\shorttitle{Observations of Microlensed Images with Dual-field Interferometry}
\shortauthors{P. Mr\'oz et al.}

\begin{document}

\title{Observations of Microlensed Images with Dual-field Interferometry: On-sky Demonstration and Prospects}

\correspondingauthor{Przemek Mr\'oz}
\email{pmroz@astrouw.edu.pl}

\author[0000-0001-7016-1692]{Przemek Mr\'oz}
\affiliation{Astronomical Observatory, University of Warsaw, Al. Ujazdowskie 4, 00-478 Warszawa, Poland}

\author[0000-0002-1027-0990]{Subo Dong}
\affiliation{Department of Astronomy, School of Physics, Peking University, 5 Yiheyuan Road, Haidian District, Beijing 100871, People's Republic of China}
\affiliation{Kavli Institute of Astronomy and Astrophysics, Peking University, 5 Yiheyuan Road, Haidian District, Beijing 100871, People's Republic of China}
\affiliation{National Astronomical Observatories, Chinese Academy of Science, 20A Datun Road, Chaoyang District, Beijing 100101, People's Republic of China}

\author[0000-0003-2125-0183]{Antoine M\'erand}
\affiliation{European Southern Observatory, Karl-Schwarzschild-Stra{\ss}e 2, D-85748 Garching, Germany}

\author[0000-0002-4569-9009]{Jinyi Shangguan}
\affiliation{Max Planck Institute for Extraterrestrial Physics, Giessenbachstra{\ss}e 1, D-85748 Garching, Germany}

\author{Julien Woillez}
\affiliation{European Southern Observatory, Karl-Schwarzschild-Stra{\ss}e 2, D-85748 Garching, Germany}

\author{Andrew Gould}
\affiliation{Department of Astronomy, Ohio State University, 140 West 18th Ave., Columbus, OH 43210, USA}
\affiliation{Max Planck Institute for Astronomy, K\"onigstuhl 17, 69117 Heidelberg, Germany}

\author[0000-0001-5207-5619]{Andrzej Udalski}
\affiliation{Astronomical Observatory, University of Warsaw, Al. Ujazdowskie 4, 00-478 Warszawa, Poland}

\author{Frank Eisenhauer}
\affiliation{Max Planck Institute for Extraterrestrial Physics, Giessenbachstra{\ss}e 1, D-85748 Garching, Germany}

\author[0000-0001-9823-2907]{Yoon-Hyun Ryu} 
\affiliation{Korea Astronomy and Space Science Institute, Daejeon 34055, Republic of Korea}

\author[0009-0007-5754-6206]{Zexuan Wu}
\affiliation{Department of Astronomy, School of Physics, Peking University, 5 Yiheyuan Road, Haidian District, Beijing 100871, People's Republic of China}
\affiliation{Kavli Institute of Astronomy and Astrophysics, Peking University, 5 Yiheyuan Road, Haidian District, Beijing 100871, People's Republic of China}

\author{Zhuokai Liu}
\affiliation{Department of Astronomy, School of Physics, Peking University, 5 Yiheyuan Road, Haidian District, Beijing 100871, People's Republic of China}
\affiliation{Kavli Institute of Astronomy and Astrophysics, Peking University, 5 Yiheyuan Road, Haidian District, Beijing 100871, People's Republic of China}

\author[0000-0003-0626-8465]{Hongjing Yang}
\affiliation{Department of Astronomy, Tsinghua University, Beijing 100084, China}

\nocollaboration{12}

\author{Guillaume Bourdarot}
\affiliation{Max Planck Institute for Extraterrestrial Physics, Giessenbachstra{\ss}e 1, D-85748 Garching, Germany}

\author{Denis Defr{\`e}re}
\affiliation{Institute of Astronomy, KU Leuven, Celestijnenlaan 200D, 3001, Leuven, Belgium}

\author{Antonia Drescher}
\affiliation{Max Planck Institute for Extraterrestrial Physics, Giessenbachstra{\ss}e 1, D-85748 Garching, Germany}

\author{Maximilian Fabricius}
\affiliation{Max Planck Institute for Extraterrestrial Physics, Giessenbachstra{\ss}e 1, D-85748 Garching, Germany}

\author{Paulo Garcia}
\affiliation{CENTRA -- Centro de Astrof{\'i}sica e Gravita\c{c}\~{a}o, IST, Universidade de Lisboa, 1049-001 Lisboa, Portugal}
\affiliation{Faculdade de Engenharia, Universidade do Porto, Rua Dr Roberto Frias, 4200-465 Porto, Portugal}

\author{Reinhard Genzel}
\affiliation{Max Planck Institute for Extraterrestrial Physics, Giessenbachstra{\ss}e 1, D-85748 Garching, Germany}

\author{Stefan Gillessen}
\affiliation{Max Planck Institute for Extraterrestrial Physics, Giessenbachstra{\ss}e 1, D-85748 Garching, Germany}

\author{Sebastian F. H\"onig}
\affiliation{School of Physics \& Astronomy, University of Southampton, Southampton, SO17 1BJ, UK}

\author{Laura Kreidberg}
\affiliation{Max Planck Institute for Astronomy, K\"onigstuhl 17, 69117 Heidelberg, Germany}

\author{Jean-Baptiste Le Bouquin}
\affiliation{Univ. Grenoble Alpes, CNRS, IPAG, 38000 Grenoble, France}

\author{Dieter Lutz}
\affiliation{Max Planck Institute for Extraterrestrial Physics, Giessenbachstra{\ss}e 1, D-85748 Garching, Germany}

\author{Florentin Millour}
\affiliation{Universit{\'e} C{\^o}te d’Azur, Observatoire de la C{\^o}te d’Azur, CNRS, Laboratoire Lagrange, Nice, France}

\author{Thomas Ott}
\affiliation{Max Planck Institute for Extraterrestrial Physics, Giessenbachstra{\ss}e 1, D-85748 Garching, Germany}

\author{Thibaut Paumard}
\affiliation{LESIA, Observatoire de Paris, Universit{\'e} PSL, Sorbonne Universit{\'e}, Universit{\'e} Paris Cit{\'e}, CNRS, 5 place Jules Janssen, 92195 Meudon, France}

\author{Jonas Sauter}
\affiliation{Max Planck Institute for Extraterrestrial Physics, Giessenbachstra{\ss}e 1, D-85748 Garching, Germany}
\affiliation{Max Planck Institute for Astronomy, K\"onigstuhl 17, 69117 Heidelberg, Germany}

\author{T. Taro Shimizu}
\affiliation{Max Planck Institute for Extraterrestrial Physics, Giessenbachstra{\ss}e 1, D-85748 Garching, Germany}

\author{Christian Straubmeier}
\affiliation{1st Institute of Physics, University of Cologne, Z{\"u}lpicher Stra{\ss}e 77, 50937 Cologne, Germany}

\author{Matthias Subroweit}
\affiliation{1st Institute of Physics, University of Cologne, Z{\"u}lpicher Stra{\ss}e 77, 50937 Cologne, Germany}

\author{Felix Widmann}
\affiliation{Max Planck Institute for Extraterrestrial Physics, Giessenbachstra{\ss}e 1, D-85748 Garching, Germany}

\collaboration{19}{(The GRAVITY+ Collaboration)}

\author[0000-0002-0548-8995]{Micha\l{} K. Szyma\'nski}
\affiliation{Astronomical Observatory, University of Warsaw, Al. Ujazdowskie 4, 00-478 Warszawa, Poland}

\author[0000-0002-7777-0842]{Igor Soszy\'nski}
\affiliation{Astronomical Observatory, University of Warsaw, Al. Ujazdowskie 4, 00-478 Warszawa, Poland}

\author[0000-0002-2339-5899]{Pawe\l{} Pietrukowicz}
\affiliation{Astronomical Observatory, University of Warsaw, Al. Ujazdowskie 4, 00-478 Warszawa, Poland}

\author[0000-0003-4084-880X]{Szymon Koz\l{}owski}
\affiliation{Astronomical Observatory, University of Warsaw, Al. Ujazdowskie 4, 00-478 Warszawa, Poland}

\author[0000-0002-9245-6368]{Rados\l{}aw Poleski}
\affiliation{Astronomical Observatory, University of Warsaw, Al. Ujazdowskie 4, 00-478 Warszawa, Poland}

\author[0000-0002-2335-1730]{Jan Skowron}
\affiliation{Astronomical Observatory, University of Warsaw, Al. Ujazdowskie 4, 00-478 Warszawa, Poland}

\author[0000-0001-6364-408X]{Krzysztof Ulaczyk}
\affiliation{Department of Physics, University of Warwick, Coventry CV4 7 AL, UK}
\affiliation{Astronomical Observatory, University of Warsaw, Al. Ujazdowskie 4, 00-478 Warszawa, Poland}

\author[0000-0002-1650-1518]{Mariusz Gromadzki}
\affiliation{Astronomical Observatory, University of Warsaw, Al. Ujazdowskie 4, 00-478 Warszawa, Poland}

\author[0000-0002-9326-9329]{Krzysztof Rybicki}
\affiliation{Department of Particle Physics and Astrophysics, Weizmann Institute of Science, Rehovot 76100, Israel}
\affiliation{Astronomical Observatory, University of Warsaw, Al. Ujazdowskie 4, 00-478 Warszawa, Poland}

\author[0000-0002-6212-7221]{Patryk Iwanek}
\affiliation{Astronomical Observatory, University of Warsaw, Al. Ujazdowskie 4, 00-478 Warszawa, Poland}

\author[0000-0002-3051-274X]{Marcin Wrona}
\affiliation{Department of Astrophysics and Planetary Sciences, Villanova University, 800 Lancaster Ave., Villanova, PA 19085, USA}
\affiliation{Astronomical Observatory, University of Warsaw, Al. Ujazdowskie 4, 00-478 Warszawa, Poland}

\author{Mateusz J. Mr{\'o}z}
\affiliation{Astronomical Observatory, University of Warsaw, Al. Ujazdowskie 4, 00-478 Warszawa, Poland}

\collaboration{12}{(The OGLE Collaboration)}

\author[0000-0003-3316-4012]{Michael D. Albrow}
\affiliation{University of Canterbury, School of Physical and Chemical Sciences, Private Bag 4800, Christchurch 8020, New Zealand}

\author[0000-0001-6285-4528]{Sun-Ju Chung}
\affiliation{Korea Astronomy and Space Science Institute, Daejeon 34055, Republic of Korea}

\author[0000-0002-2641-9964]{Cheongho Han}
\affiliation{Department of Physics, Chungbuk National University, Cheongju 28644, Republic of Korea}

\author[0000-0002-9241-4117]{Kyu-Ha Hwang}
\affiliation{Korea Astronomy and Space Science Institute, Daejeon 34055, Republic of Korea}

\author[0000-0002-0314-6000]{Youn Kil Jung}
\affiliation{Korea Astronomy and Space Science Institute, Daejeon 34055, Republic of Korea}
\affiliation{National University of Science and Technology (UST), Daejeon 34113, Republic of Korea}

\author[0000-0002-4355-9838]{In-Gu Shin}
\affiliation{Center for Astrophysics $|$ Harvard \& Smithsonian, 60 Garden St.,Cambridge, MA 02138, USA}

\author[0000-0003-1525-5041]{Yossi Shvartzvald}
\affiliation{Department of Particle Physics and Astrophysics, Weizmann Institute of Science, Rehovot 7610001, Israel}

\author[0000-0001-9481-7123]{Jennifer C. Yee}
\affiliation{Center for Astrophysics $|$ Harvard \& Smithsonian, 60 Garden St.,Cambridge, MA 02138, USA}


\author[0000-0001-6000-3463]{Weicheng Zang}
\affiliation{Center for Astrophysics $|$ Harvard \& Smithsonian, 60 Garden St.,Cambridge, MA 02138, USA}


\author[0000-0002-7511-2950]{Sang-Mok Cha} 
\affiliation{Korea Astronomy and Space Science Institute, Daejeon 34055, Republic of Korea}
\affiliation{School of Space Research, Kyung Hee University, Yongin, Kyeonggi 17104, Republic of Korea} 

\author{Dong-Jin Kim}
\affiliation{Korea Astronomy and Space Science Institute, Daejeon 34055, Republic of Korea}

\author[0000-0003-0562-5643]{Seung-Lee Kim} 
\affiliation{Korea Astronomy and Space Science Institute, Daejeon 34055, Republic of Korea}

\author[0000-0003-0043-3925]{Chung-Uk Lee}
\affiliation{Korea Astronomy and Space Science Institute, Daejeon 34055, Republic of Korea}

\author[0009-0000-5737-0908]{Dong-Joo Lee} 
\affiliation{Korea Astronomy and Space Science Institute, Daejeon 34055, Republic of Korea}

\author[0000-0001-7594-8072]{Yongseok Lee} 
\affiliation{Korea Astronomy and Space Science Institute, Daejeon 34055, Republic of Korea}
\affiliation{School of Space Research, Kyung Hee University, Yongin, Kyeonggi 17104, Republic of Korea}

\author[0000-0002-6982-7722]{Byeong-Gon Park}
\affiliation{Korea Astronomy and Space Science Institute, Daejeon 34055, Republic of Korea}

\author[0000-0003-1435-3053]{Richard W. Pogge} 
\affiliation{Department of Astronomy, Ohio State University, 140 West 18th Ave., Columbus, OH  43210, USA}
\affiliation{Center for Cosmology and AstroParticle Physics, Ohio State University, 191 West Woodruff Ave., Columbus, OH 43210, USA}

\collaboration{17}{(The KMTNet Collaboration)}

\begin{abstract}
Interferometric observations of gravitational microlensing events offer an opportunity for precise, efficient, and direct mass and distance measurements of lensing objects, especially those of isolated neutron stars and black holes. However, such observations have previously been possible for only a handful of extremely bright events. The recent development of a dual-field interferometer, GRAVITY Wide, has made it possible to reach out to significantly fainter objects and increase the pool of microlensing events amenable to interferometric observations by two orders of magnitude. Here, we present the first successful observation of a microlensing event with GRAVITY Wide and the resolution of microlensed images in the event OGLE-2023-BLG-0061/KMT-2023-BLG-0496. We measure the angular Einstein radius of the lens with subpercent precision, $\thetaE = 1.280 \pm 0.009$~mas. Combined with the microlensing parallax detected from the event light curve, the mass and distance to the lens are found to be $0.472 \pm 0.012\,M_{\odot}$ and $1.81 \pm 0.05$\,kpc, respectively. We present the procedure for the selection of targets for interferometric observations and discuss possible systematic effects affecting GRAVITY Wide data. This detection demonstrates the capabilities of the new instrument, and it opens up completely new possibilities for the follow-up of microlensing events and future routine discoveries of isolated neutron stars and black holes.
\end{abstract}


\section{Introduction} \label{sec:intro}

In his seminal paper on gravitational microlensing, \citet{einstein1936} realized that the angular separation between the two images of the source star created by a Galactic gravitational lens would be well below the resolution limits of contemporary telescopes. That led him to believe that observing microlensed images would be, in practice, impossible. This notion persisted for many decades until the first optical interferometry facilities were developed.

In point-source point-lens microlensing events, the gravitational lens creates two images of the source star (called the major and minor image). They are separated by approximately $2\thetaE$, where $\thetaE=\sqrt{\kappa M \pirel}$ is the angular Einstein radius. Here, $M$ is the lens mass, $\pirel$ is the lens--source relative parallax, and $\kappa=8.144\,\mathrm{mas}\,M_{\odot}^{-1}$ is a constant. For typical configurations of stellar-mass microlensing events in the Milky Way, $\thetaE \sim 1\,\mathrm{mas}$, and so the expected separation between the minor and major images is usually smaller than a few milliarcseconds. Interferometric observations of microlensing events, therefore, provide a direct way to precisely measure angular Einstein radii by resolving the images (and, as a consequence, measure the masses of lensing objects).

Although a few tens of thousands of single-lens microlensing events have been discovered so far, only a few have had the angular Einstein radius measured. That was possible for only a small number of events exhibiting finite-source effects \citep{gould1994,nemi1994,witt1994}, and the probability of such measurements is biased toward low-mass (planetary-mass) lenses. Another route for $\thetaE$ measurements has involved astrometric observations \citep{hog1995,miyamoto1995,walker1995}, leading to the first detection of an isolated black hole \citep{sahu2022,lam2022,mroz2022,lam2023}. These observations, however, required long-term astrometric monitoring, which is not feasible for a large sample of events. Conversely, interferometric observations are much more efficient, as only one exposure is sufficient to resolve the microlensed images and measure the angular Einstein radius. When combined with the microlensing parallax ($\piE \equiv \pirel / \thetaE$) measurements from the light curve, they allow us to precisely determine the masses, distances, and transverse velocities of isolated objects, including neutron stars and black holes.

\citet{delplancke2001} were the first to discuss the prospects of using the European Southern Observatory (ESO) Very Large Telescope Interferometer (VLTI) to study gravitational microlensing events. They estimated that the first-generation VLTI instruments would be able to observe dozens of events every year in the mid 2000s. Yet, the first successful resolution of microlensed images did not materialize until late 2017, when \citet{dong2019} observed the microlensing event TCP J05074264+2447555 (aka Kojima-1) with the second-generation VLTI instrument GRAVITY \citep{gravity2017}.

\begin{figure}
\includegraphics[width=.5\textwidth]{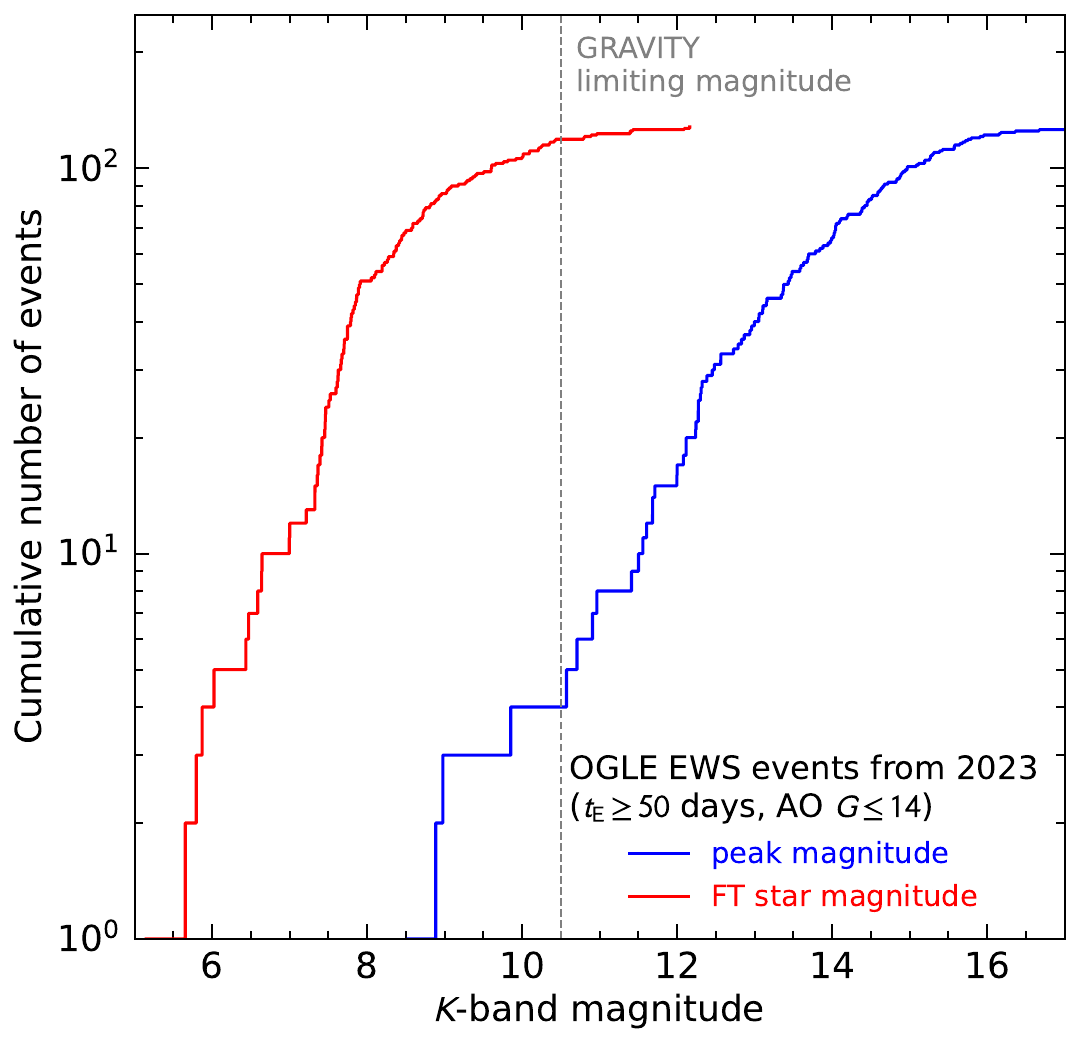}
\caption{Cumulative distribution of the expected $K$-band peak magnitudes of events detected by the OGLE EWS in 2023 (blue line). For comparison, the red line shows the cumulative distribution of the $K$-band magnitudes of the brightest star within $30''$ of the event (fringe-tracking star). Only events with $\tE \geq 50$\,days and an adaptive optics guide star brighter than $G=14$ are plotted.}
\label{fig:stats}
\end{figure}

Why did it take almost two decades to achieve this milestone? The early predictions of VLTI performance underestimated the role and impact of many practical problems encountered during system operations, such as beam stability, vibrations, and air dispersion \citep[e.g.,][]{delplancke2008}. Moreover, even when these technical problems were solved or mitigated, the performance of a ground-based interferometer is ultimately limited by atmospheric turbulence, which breaks the coherence of the wave fronts arriving at each telescope. Therefore, the observations must be carried out with exposure times ($\sim 1$\,ms) shorter than the typical wave-front-coherence timescale (on the order of 20--30\,ms) and shorter than the mechanical vibrations timescale, which dominate in the 10--1000\,Hz regime. Even for the largest modern 10 m class telescopes, this requirement sets a limiting magnitude of $K \approx 10$ for interferometric observations. Gravitational microlensing events brighter than this limit are exceedingly rare (Figure~\ref{fig:stats}). They occur at most a few times a year, and they exceed the GRAVITY limiting magnitude for a short period of time. The three events with published interferometric observations were unusually bright: TCP J05074264+2447555 \citep{dong2019,zang2020}, ASASSN-22av \citep{wu2024b}, and Gaia19bld \citep{cassan2022,rybicki2022,bachelet2022} reached $V\approx 11.5$, $g'\approx 12.5$, and $I\approx 9.0$, respectively. The former two were observed using the standard on-axis mode of GRAVITY, whose fringe tracker allows minute-long science exposures \citep{lacour2019} that significantly enhance the VLTI sensitivity \citep{eisenhauer2023}.

The dual-field interferometric observations were designed to alleviate this problem and reach fainter sources \citep[e.g.,][]{shao1992,colavita1999,delplancke2008,woillez2014}. In this observing mode, the target and a nearby bright ($K\!\lesssim\!10$) reference (fringe-tracking) star are observed at the same time. The fringe-tracking star is needed to measure the atmospheric turbulence and stabilize the optical path difference between the telescopes, providing a correction to the science target. That enables observation of the target with exposure times longer than the turbulence and vibration timescales, allowing one to reach fainter objects. The angular separation between the science object and the fringe-tracking star must be smaller than $20''\!-\!30''$, which is set by the atmospheric conditions. In 2019--2022, the GRAVITY instrument was upgraded to enable such dual-field interferometric observations \citep{gravity2022}, hence the name of this observing mode, GRAVITY Wide.

The upgraded instrument opened up an entirely new pathway for characterizing and studying microlensing events because, in the dense fields of the Galactic bulge, the probability of finding a suitable ($K \lesssim 10$, within $30''$) fringe-tracking star is as large as 50\%--70\%. According to the GRAVITY manual (Issue 110),\footnote{\url{https://www.eso.org/sci/facilities/paranal/instruments/gravity.html}} the GRAVITY Wide observations are possible for science targets as faint as $K_{\rm SC} \approx 16\mbox{--}17$. 
However, the limiting magnitude strongly depends on the atmospheric conditions and the separation between the science object and the fringe-tracking star. In addition, closure-phase observations lose the signal-to-noise ratio faster than normal visibility data as the targets become fainter.
Thus, in practice, we rarely considered observing targets fainter than $K \approx 13.5\mbox{--}14$, which still left us with plenty of candidate events to scrutinize (Figure~\ref{fig:stats}).

As soon as ESO offered regular GRAVITY Wide observations from October 2022 (``Period 110''), we initiated a program (PI: A.~M\'erand) of interferometric observations of microlensing events. Our primary science goal was to detect and measure precise masses, distances, and transverse velocities of isolated stellar remnants---neutron stars and black holes. Our observations also served as a test bed for verifying the capabilities of the new instrument and planning, executing, and analyzing interferometric follow-up observations of a large number of microlensing events. 

This paper presents the first resolution of microlensed images with GRAVITY Wide in the microlensing event OGLE-2023-BLG-0061/KMT-2023-BLG-0496.

\section{Selection of the Targets for Interferometric Follow-up} \label{sec:selection}

The VLTI observations can be carried out with either four 8.2 m Unit Telescopes (UTs) or four 1.8 m Auxiliary Telescopes (ATs). The VLTI UT observing runs are organized every month (around the full Moon) and typically last one week. During the rest of the month, observations are possible with ATs, which may be relocated to more than ten observing stations.\footnote{\url{https://www.eso.org/sci/facilities/paranal/telescopes/vlti.html}} GRAVITY Wide observations may be conducted with ATs, provided that the fringe-tracking star is brighter than $K_{\rm FT}\approx9\mbox{--}9.5$ and the science target is brighter than $K_{\rm SC}\approx13\mbox{--}14$. For UTs, the limiting magnitudes are larger---$K_{\rm FT}\approx10\mbox{--}10.5$ and $K_{\rm SC}\approx16\mbox{--}17$, respectively. As discussed above, the limiting magnitudes of science targets strongly depend on the atmospheric conditions (isoplanatic angle) and the separation of the fringe-tracking star.

We began the selection of microlensing targets about 7--10~days before the planned start of each VLTI UT run. The candidates were chosen from publicly available lists of microlensing alerts published by the Optical Gravitational Lensing Experiment (OGLE) Early Warning System\footnote{\url{https://ogle.astrouw.edu.pl/ogle4/ews/ews.html}} \citep[EWS;][]{udalski2003,udalski2015}, the Korea Microlensing Telescope Network (KMTNet) Alert System\footnote{\url{https://kmtnet.kasi.re.kr/~ulens/}} \citep{kim2018_alerts}, and the Microlensing Observations in Astrophysics (MOA) Transient Alerts.\footnote{\url{https://www.massey.ac.nz/~iabond/moa/alerts/}} In addition, we also checked the transient alerts published by all-sky surveys, such as \textit{Gaia} \citep{hodgkin2013,hodgkin2021} and the All Sky Automated Survey for SuperNovae \citep[ASAS-SN;][]{shappee2014}.

The selection of targets was based on several scientific and technical criteria. First, we selected events near or past their maximum brightness, so that the parameters describing their light curves were reasonably well measured. However, the observations had to be secured before the event faded, before the minor image became too faint, and so the contrast ratio between the microlensed images became too large. We required the contrast ratio to be smaller than 10:1, which was conservatively adopted based on our experience with the GRAVITY data. That contrast ratio corresponds to the maximum lens--source separation (in Einstein radius units) of $u_{\rm max} = 1.22$ or, equivalently, the minimum amplitude of $\Delta I_{\rm min} = 0.13$.

Because the primary scientific motivation of our project was searching for stellar remnants, which are expected to give rise to long-duration events, we selected events with Einstein timescales longer than $\tE = 50$ days. However, the nature of the lens cannot be known at the time of selecting the targets. The mass of the lens can be determined only after the interferometric data are combined with the full light curve. That usually means waiting several weeks after the interferometric observations are taken, because they are collected close to the maximum magnification.

In the next step, we checked if suitable fringe-tracking and adaptive optics reference stars were located within $30''$ of the event. For possible fringe-tracking stars, we queried the Two Micron All Sky Survey (2MASS) Point Source Catalog \citep{skrutskie2006}, while for adaptive optics guide stars, we queried \textit{Gaia} Data Release 3 \citep{gaia2016,gaia_dr3}. The guide star had to be brighter than $G \approx 14$ for the Multi Application Curvature Adaptive Optics \citep{arsenault2003} system or brighter than $K=8$ for the Coud\'e Infrared Adaptive Optics \citep{kendrew2012} system. Finally, we estimated the expected $K$-band brightness of the event during the planned observations, by assuming that the blending parameter in the $K$ band was identical to that in the $I$ band. The baseline $K$-band brightness of the event was taken from the VISTA Variables in the Via Lactea (VVV) survey \citep{minniti2010}, the United Kingdom Infra-Red Telescope (UKIRT) Galactic Plane Survey \citep{lawrence2007,lucas2008}, or 2MASS \citep{skrutskie2006}. We also examined the estimated $K$-band brightness of the source by using the best-fit $I$-band source fluxes as cross-checks. For such estimates, we assumed dereddened color $(I-K)_{0,\rm giant}=1.4$ and $(I-K)_{0,\rm dwarf}=1.0$ for sources roughly classified using the extinction-corrected source magnitudes as giants ($I_0 < 16.5$) and dwarfs ($I_0 > 16.5$), respectively.

Candidate targets were selected independently by two teams (P.M. and S.D.) and subsequently investigated in more detail. In particular, we paid special attention to the microlensing parallax measurements, which are necessary for the lens mass determination. Because the value of the microlensing parallax is inversely proportional to the square root of the lens mass, we required it to be consistent with zero (or close to zero) during the trigger. We also ran light-curve simulations to ensure that the parallax would be precisely measured (or constrained) by the end of the observing season. We did not consider events with variable source stars, because the variability may affect the microlensing parallax measurements.

The blue solid line in Figure~\ref{fig:stats} shows the cumulative distribution of the expected $K$-band peak magnitudes of the events detected by the OGLE EWS in 2023. Only events with relatively long timescales ($\tE \geq 50$\,days) and suitable adaptive optics guide stars ($G \leq 14$) within $30''$ are presented. Only a few events were bright enough ($K<10.5$) for standard GRAVITY on-axis observations. In contrast, the solid red line in Figure~\ref{fig:stats} shows the distribution of the $K$-band magnitudes of the brightest star within $30''$ of the event (which may serve as a fringe-tracking star). Nearly 100 events could have been considered for GRAVITY Wide observations.

\section{Data} \label{sec:data}

The detection of the microlensing event OGLE-2023-BLG-0061 was announced by the OGLE EWS \citep{udalski2003,udalski2015} on 2023 March 13.60 UT ($\mathrm{HJD'}\equiv \mathrm{HJD}-2460000=17.10$). It was independently identified by the KMTNet Alert System \citep{kim2016,kim2018_alerts} on 2023 April 20, and it was designated KMT-2023-BLG-0496. The event occurred on a bright red clump star ($I=16.389 \pm 0.001$, $V-I=2.26 \pm 0.02$) with equatorial coordinates (R.A., Decl.)$_{\rm J2000}$ = (\ra{17}{43}{04}{01}, \dec{-35}{15}{32}{3}). According to data from the VVV survey, the source star had $K=13.515 \pm 0.011$ \citep{minniti2010}.

\subsection{Photometric Data}

OGLE operates the 1.3~m Warsaw Telescope located at Las Campanas Observatory, Chile. The telescope is equipped with a mosaic camera covering a field of view of 1.4\,deg$^2$ with a pixel scale of $0.26\,\mathrm{arcsec}\,\mathrm{pixel}^{-1}$. The event has been observed by the OGLE-IV survey since 2010. However, in this paper, we analyze the OGLE data collected from 2016 through 2023, because earlier observations do not contribute to constraining the parameters of the model. Observations were reduced using the OGLE-IV data reduction pipeline \citep{udalski2015}, which employs a custom implementation of the Difference Image Analysis (DIA) method \citep{wozniak2000}.

KMTNet uses three 1.6 m telescopes located at the Cerro Tololo Inter-American Observatory (KMTC; Chile), the South African Astronomical Observatory (KMTS; South Africa), and the Siding Spring Observatory (KMTA; Australia). Each of the KMTNet telescopes is equipped with a camera with a 4\,deg$^2$ field of view and a pixel scale of $0.40\,\mathrm{arcsec}\,\mathrm{pixel}^{-1}$. The analyzed KMTNet observations cover the years 2021--2023. However, because of saturation, we deleted data points near the peak of the event. The KMTNet photometric data were reduced with the tender-loving-care (TLC) DIA-based pipeline \citep{yang2024}, which was developed from pySIS \citep{albrow2009}. The vast majority of OGLE and KMTNet images were taken in the $I$-band filter, with additional $V$-band observations to characterize the color of the source star.

Some additional observations were taken in the $R'$ band with the 0.18 m Newtonian telescope (CHI-18; \citealt{wu2024}), located at the El Sauce Observatory in Chile, to cover the peak of the event. The CHI-18 images were reduced with the TLC pipeline. The original observations were taken at a 2 minutes cadence. Because the event did not exhibit variability on such short timescales, we binned the CHI-18 data into 1 hr long bins. 

\begin{deluxetable*}{ccccc}
\tablecaption{Log of VLTI Observations\label{tab:obs}}
\tablehead{\colhead{Epoch} & \colhead{Time} & \colhead{HJD$'$} & \colhead{Resolution} & \colhead{Exposure Time}}
\startdata
1a & 2023-07-29 02:59:45 UT & 154.625 & Medium & $4 \times 100$ s\\
1b & 2023-07-29 03:21:42 UT & 154.640 & Low    & $12 \times 30$ s\\
2a & 2023-09-28 23:30:05 UT & 216.479 & Low    & $12 \times 30$ s\\
2b & 2023-09-28 23:36:41 UT & 216.484 & Low    & $12 \times 30$ s\\
2c & 2023-09-28 23:43:20 UT & 216.488 & Low    & $12 \times 30$ s\\
2d & 2023-09-28 23:57:23 UT & 216.498 & Low    & $12 \times 30$ s\\
2e & 2023-09-29 00:03:56 UT & 216.503 & Low    & $12 \times 30$ s\\
2f & 2023-09-29 00:10:29 UT & 216.507 & Low    & $12 \times 30$ s\\
\enddata
\end{deluxetable*}

\begin{figure*}
\includegraphics[width=.5\textwidth]{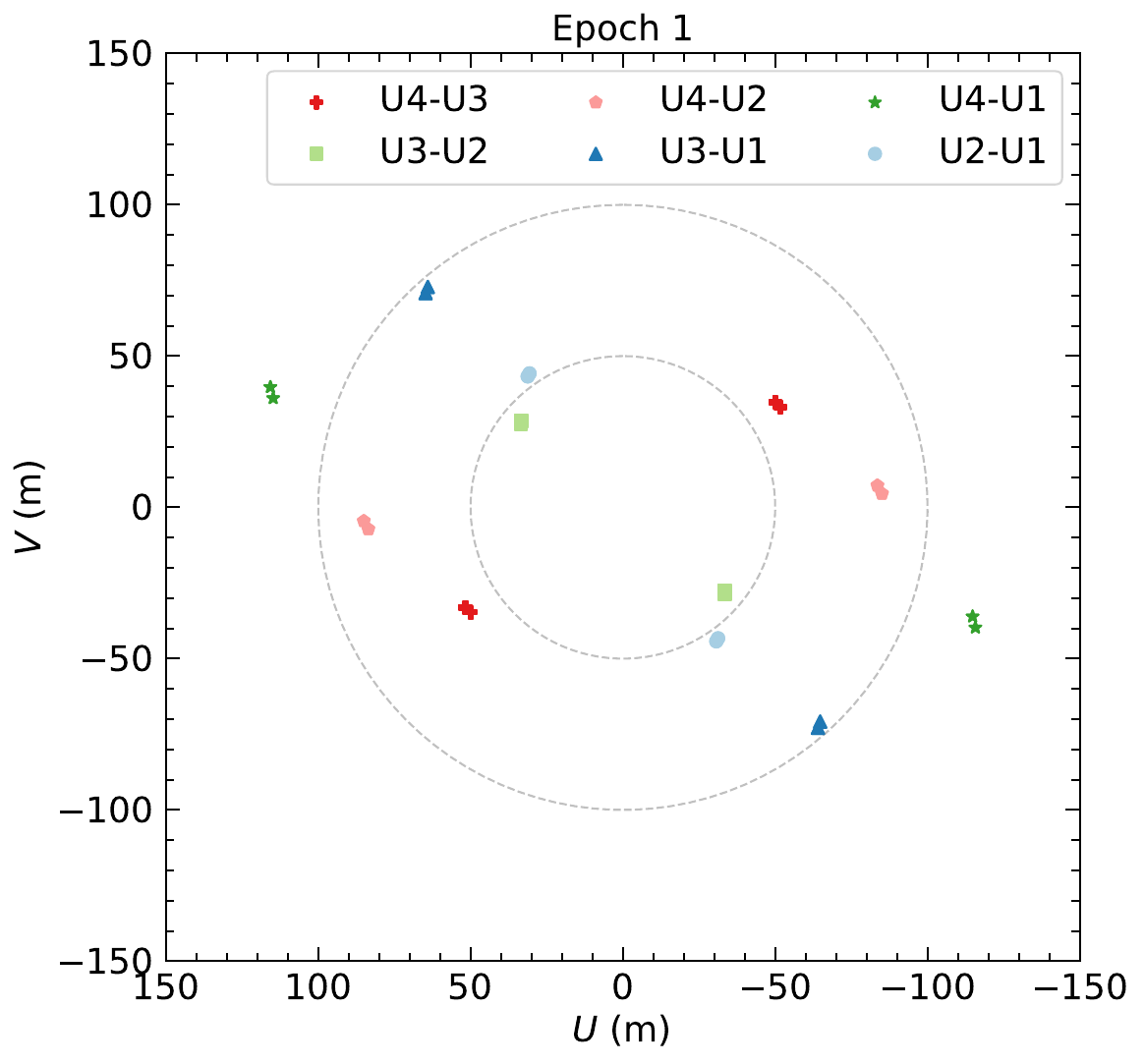}
\includegraphics[width=.5\textwidth]{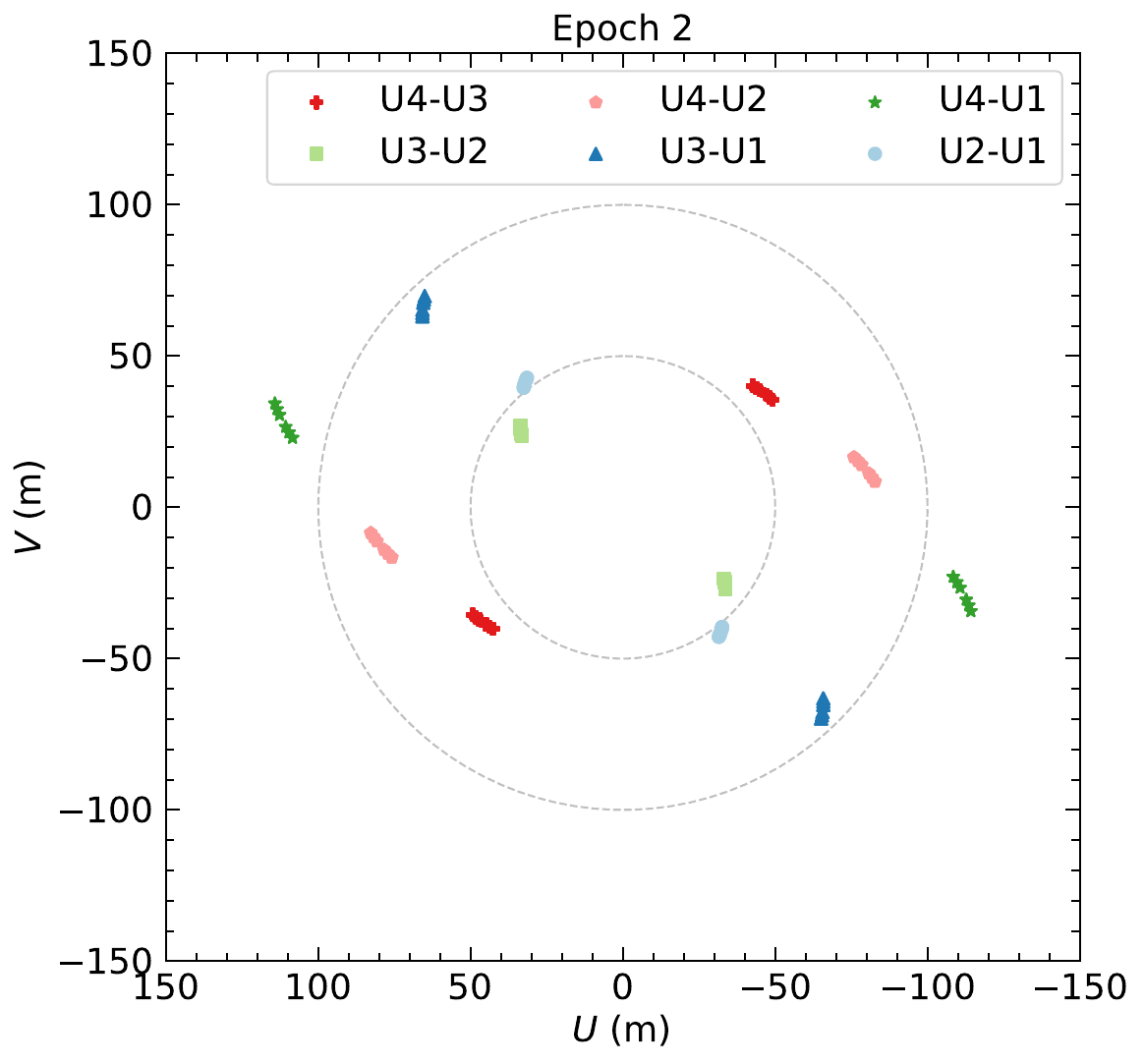}
\caption{UV plane coverage of epoch 1 (left panel) and epoch 2 (right panel) GRAVITY observations of OGLE-2023-BLG-0061/KMT-2023-BLG-0496.}
\label{fig:uv}
\end{figure*}

\subsection{VLTI Data}

OGLE-2023-BLG-0061/KMT-2023-BLG-0496 was considered a promising candidate for VLTI observations on 2023 July 20. Although the event was still before the peak, the available data predicted a relatively long timescale ($\tE \approx 100$ days). At the same time, the optical brightness indicated that the microlens parallax should be robustly constrained. A nearby ($13.2''$) bright ($G=13.7$, $K=9.8$) star, 2MASS 17430426-3515195, could serve as a fringe-tracking and adaptive optics guide star. The first GRAVITY Wide observations were secured with UTs on 2023 July 29 (``epoch 1''). We obtained four sets of medium-resolution observations ($R=\lambda/\Delta\lambda\approx 500$) and one set of low-resolution data ($R\approx 22$), a new mode never tried before with GRAVITY Wide. The first three medium-resolution exposures were of low quality, and the resulting closure phases were very noisy. We therefore decided not to use them in the subsequent analysis.  The preliminary modeling carried out at that time indicated that the medium- and low-resolution data were inconsistent with each other.

We thus attempted to collect additional VLTI data to study the magnitude of the possible systematics affecting the low-resolution data. We triggered on-axis GRAVITY observations with ATs on 2023 August 23, when the event was approaching the peak (at magnification $\sim 90$), but the data could not be collected. The next (successful) attempt to gather additional VLTI UT observations took place on 2023 September 29 (``epoch 2''), when we secured six low-resolution exposures with GRAVITY Wide. Table~\ref{tab:obs} presents the log of VLTI observations and Figure~\ref{fig:uv} shows the coverage of the UV plane during both VLTI epochs. Each low-resolution observation consisted of twelve 30\,s exposures; the medium-resolution observation consisted of four 100\,s exposures. For each night, a bright-star pair was observed to center the science fringe with the GRAVITY differential delay line before the microlens observation. We did not adjust the science fringe when the telescope was moved to the science target.

The data were reduced with the standard GRAVITY pipeline (version 1.4.2). We first used the Python script run\_gravi\_reduced.py to reduce the raw data and apply the pixel-to-visibility matrix (P2VM). The default options were used, except that we adopted --gravity\_vis.output-phase-sc=SELF\_VISPHI to calculate the internal differential phase between each spectral channel and --gravity\_vis.opd-pupil-stddev-max-sc=9999 to ignore the poor pupil measurements in the acquisition camera, which do not affect our closure-phase measurements. The pipeline performed the bias and sky subtraction, flat-fielding, wavelength calibration, and spectral extraction. The application of the P2VM converts the pixel detector counts into complex visibilities, taking into account all instrumental effects, including relative throughput, coherence, phase shift, and crosstalk. The dark, bad-pixel, flat-field, wavelength calibration, and P2VM matrix data were reduced from the daily calibration data obtained close in time to our observations.  We then used run\_gravi\_trend.py to calibrate the closure-phase data. For epoch~1, we used the star pair to center the science fringe and calibrate the medium-resolution data. Unfortunately, we did not have calibrator data observed in the low-resolution data on the same night. For epoch~2, we used a bright-star pair observed after the microlens observation for the calibration. The following analyses are based on the closure-phase data from the calibrated medium-resolution and uncalibrated low-resolution data from epoch~1 and the calibrated low-resolution data from epoch~2. In this way, we use all the data with good quality. Meanwhile, when modeling the low-resolution data from epoch~2, we found that the calibration makes little difference to the closure phase.

\section{Light-curve Model}
\label{sec:light_curve_model}

\begin{figure}
\centering
\includegraphics[width=.5\textwidth]{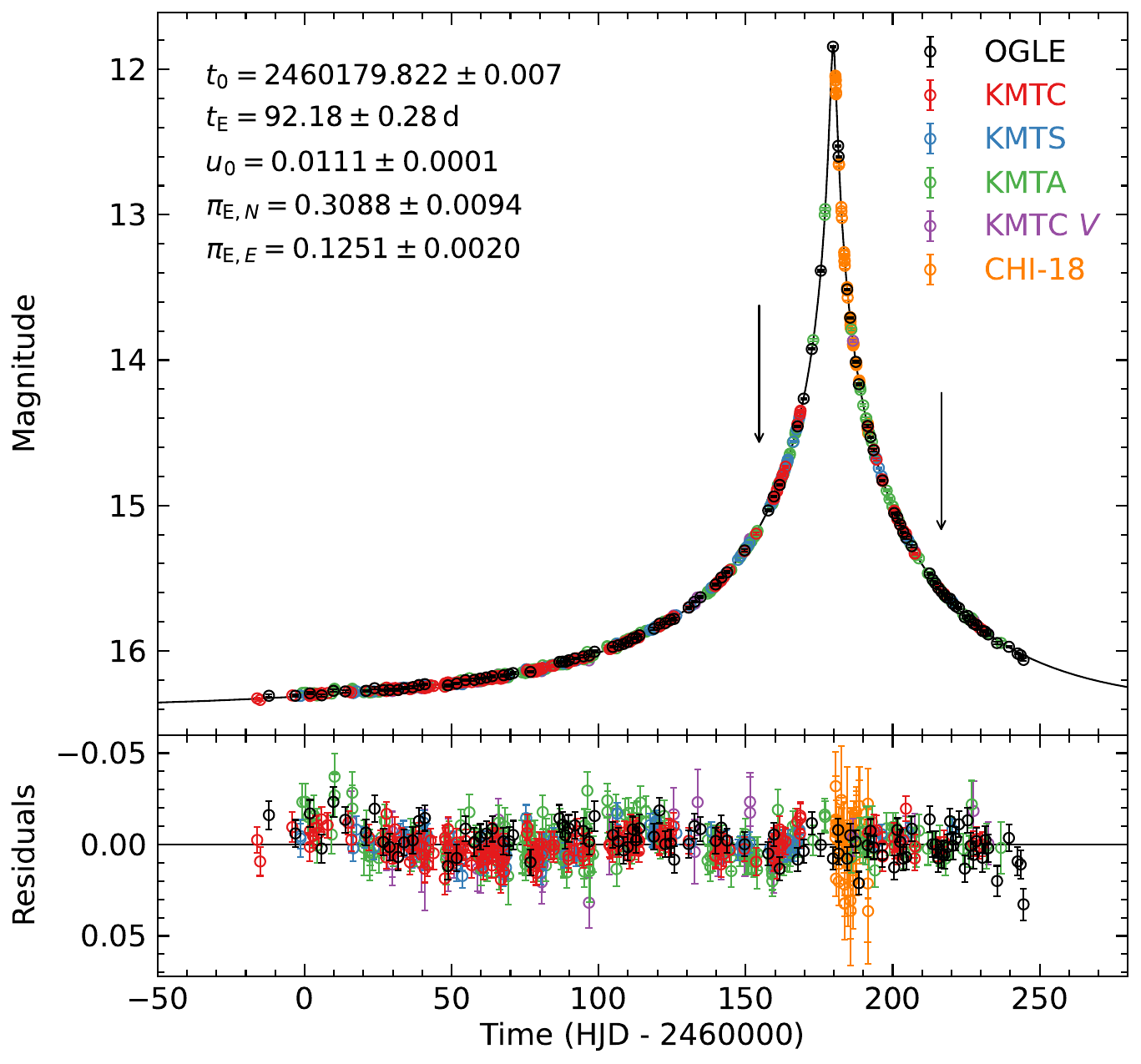}
\caption{Light curve of OGLE-2023-BLG-0061/KMT-2023-BLG-0496. The black line marks the best-fit model with $u_0>0$. Arrows mark the two epochs of VLTI observations.}
\label{fig:phot}
\end{figure}

The light curve of the event can be well fitted by a standard point-source point-lens model with the annual parallax effect (which is caused by the orbital motion of the Earth). This model has five free parameters: the time of closest approach between the lens and the source $t_0$, their minimum separation (in Einstein radius units) $u_0$, the Einstein radius crossing timescale $\tE$, and the northern and eastern components of the microlensing parallax vector $\piEvec = (\piEN,\piEE)$. The latter is a vector quantity whose direction is parallel to the direction of the relative lens--source proper motion $\boldsymbol{\mu}_{\rm rel}$. The magnification is calculated using the formula
\begin{equation}
A(t) = \frac{u(t)^2+2}{u(t)\sqrt{u(t)^2+4}},
\end{equation}
where $u(t)=\sqrt{\tau (t)^2+\beta (t)^2}$. The latter quantity is evaluated in the geocentric frame that is moving with a velocity equal to the Earth's velocity at $t_{0,\rm{par}}=2460179$ \citep{gould2004} in which
\begin{equation}
\tau(t) = \frac{t-t_0}{\tE}+\delta\tau(t),\quad \beta(t) = u_0+\delta\beta(t)
\end{equation}
and
\begin{equation}
(\delta\tau,\delta\beta)=(\piEvec \cdot \Delta s,\piEvec\times \Delta s),
\end{equation}
where $\Delta s$ is the projected position of the Sun. Two models with different signs of $u_0$ are possible; both models are almost perfectly degenerate, with the $u_0<0$ model being preferred by only $\Delta\chi^2=3.1$. The best-fit model (with $u_0>0$) is presented in Figure~\ref{fig:phot}. The best-fit parameters and their uncertainties are reported in Table~\ref{tab:phot_params}. In Table~\ref{tab:phot_params}, we also report the best-fit source magnitude $I_{\rm s}$, baseline magnitude $I_0$, and dimensionless blending parameter $f_{\rm s}$.

\begin{deluxetable}{lrr}
\caption{Best-fit Parameters of the Light-curve Model\label{tab:phot_params}}
\tablehead{\colhead{Parameter} & \colhead{$u_0>0$} & \colhead{$u_0<0$}}
\startdata
$t_0$ (HJD$'$)        & $179.8223 \pm 0.0066$ & $179.8103 \pm 0.0063$ \\
$u_0$               & $0.01112 \pm 0.00008$ & $-0.01112 \pm 0.00008$\\
$\tE$ (days)         & $92.18 \pm 0.28$      & $91.93 \pm 0.28$      \\
$\piEN$             & $0.3088 \pm 0.0094$   & $0.3222 \pm 0.0099$   \\
$\piEE$             & $0.1251 \pm 0.0020$   & $0.1245 \pm 0.0021$   \\
\hline
$\chi^2$/dof     & $2093.7/2069$         & $2090.6/2069$         \\
\hline
$\piE$              & $0.3330 \pm 0.0084$   & $0.3454 \pm 0.0090$   \\
$\Phi_{\pi}$ (deg)  & $22.07 \pm 0.81$      & $21.13 \pm 0.81$      \\
$f_{\rm s}$ (OGLE)  & $0.7392 \pm 0.0030$   & $0.7353 \pm 0.0029$   \\
$I_{\rm s}$ (OGLE)  & $16.717 \pm 0.005$    & $16.723 \pm 0.005$    \\
$I_0$ (OGLE)        & $16.389 \pm 0.001$    & $16.389 \pm 0.001$    \\
$u(1)$              & $0.2640 \pm 0.0008$   & $0.2629 \pm 0.0008$   \\
$\eta (1)$          & $0.5907 \pm 0.0009$   & $0.5920 \pm 0.0009$   \\
$\phi (1)$ (deg)    & $173.11 \pm 0.15$     & $177.73 \pm 0.15$     \\
$\mathrm{PA}(1)$ (deg) & $-164.83 \pm 0.95$    & $-156.60 \pm 0.68$    \\
$u(2)$              & $0.4295 \pm 0.0015$   & $0.4279 \pm 0.0015$   \\
$\eta (2)$          & $0.4264 \pm 0.0013$   & $0.4277 \pm 0.0013$   \\
$\phi (2)$ (deg)    & $9.46 \pm 0.24$       & $6.84 \pm 0.26$       \\
$\mathrm{PA}(2)$ (deg) & $31.52 \pm 0.58$      & $14.29 \pm 1.06$      \\
\enddata
\tablecomments{Parameters denoted by (1) and (2) were calculated for the VLTI epochs~1 and~2, respectively.}
\end{deluxetable}

\begin{figure}
\centering
\includegraphics[width=.5\textwidth]{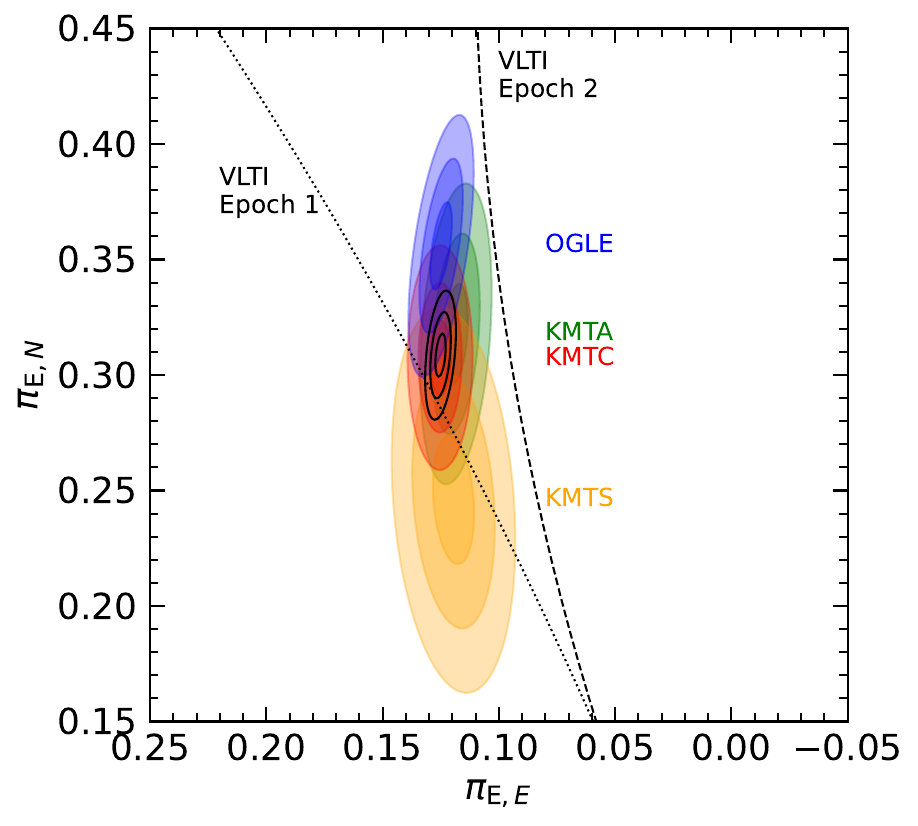}
\caption{Constraints on the microlensing parallax derived from different photometric data sets (blue---OGLE; red---KMTC; orange---KMTS; and green---KMTA) and all light-curve data (black contours). The dotted and dashed lines mark the constraints from the two epochs of VLTI data, respectively.}
\label{fig:parallaxes}
\end{figure}

Figure~\ref{fig:parallaxes} shows the constraints on the microlensing parallax vector derived using data from different observatories. The blue, red, green, and orange contours mark the constraints from OGLE, KMTC, KMTA, and KMTS data, respectively, whereas the solid black contours mark the best-fit model to all data. 
The eastern component of $\piEvec$, which is parallel to the projected acceleration of the Sun, is relatively well measured in all data sets. However, the northern component $\piEN$ (perpendicular to the projected acceleration of the Sun) has a considerably larger uncertainty. The northern component is also more susceptible to noise in the data \citep{gould1994_1dparallax,smith2003,gould2004}; hence, slightly different values of $\piEN$ are determined using different data sets. The best-fit parameters for individual data sets are reported in Tables~\ref{tab:phot_params_all} and~\ref{tab:phot_params_all2}.

\begin{deluxetable*}{lrrrr}
\caption{Best-fit Parameters of the Light-curve Model $(u_0>0)$ for Individual Data Sets\label{tab:phot_params_all}}
\tablehead{\colhead{Parameter} & \colhead{OGLE} & \colhead{KMTC} & \colhead{KMTS} & \colhead{KMTA}}
\startdata
$t_0$ (HJD')        & $179.8290 \pm 0.0082$ & $179.8135 \pm 0.0313$ & $179.8279 \pm 0.0401$ & $179.8406 \pm 0.0189$ \\
$u_0$               & $0.01099 \pm 0.00011$ & $0.00745 \pm 0.00432$ & $0.00841 \pm 0.00472$ & $0.00806 \pm 0.00112$ \\
$\tE$ (days)            & $93.57 \pm 0.53$      & $91.32 \pm 0.70$      & $93.89 \pm 1.43$      & $93.97 \pm 0.74$      \\
$\piEN$             & $0.3559 \pm 0.0199$   & $0.3076 \pm 0.0162$   & $0.2463 \pm 0.0282$   & $0.3182 \pm 0.0220$   \\
$\piEE$             & $0.1248 \pm 0.0039$   & $0.1252 \pm 0.0047$   & $0.1194 \pm 0.0088$   & $0.1184 \pm 0.0050$   \\
\hline
$\chi^2$/dof     & $542.9/541$           & $600.8/632$           & $329.3/329$           & $417.2/417$           \\
\hline
$\piE$              & $0.3772 \pm 0.0178$   & $0.3321 \pm 0.0150$   & $0.2738 \pm 0.0263$   & $0.3395 \pm 0.0200$   \\
$\Phi_{\pi}$ (deg)  & $19.33 \pm 1.52$      & $22.15 \pm 1.37$      & $25.90 \pm 2.93$      & $20.41 \pm 1.78$      \\
$u(1)$              & $0.2601 \pm 0.0014$   & $0.2661 \pm 0.0022$   & $0.2593 \pm 0.0041$   & $0.2592 \pm 0.0019$   \\
$\eta (1)$          & $0.5953 \pm 0.0016$   & $0.5882 \pm 0.0026$   & $0.5962 \pm 0.0048$   & $0.5964 \pm 0.0023$   \\
$\phi (1)$ (deg)    & $172.30 \pm 0.32$     & $173.98 \pm 0.96$     & $174.56 \pm 1.15$     & $173.50 \pm 0.45$     \\
$\mathrm{PA}(1)$ (deg) & $-168.37 \pm 1.83$ & $-163.90 \pm 1.75$    & $-159.58 \pm 3.50$    & $-166.08 \pm 2.17$    \\
$u(2)$              & $0.4254 \pm 0.0025$   & $0.4328 \pm 0.0039$   & $0.4354 \pm 0.0066$   & $0.4204 \pm 0.0039$   \\
$\eta (2)$          & $0.4298 \pm 0.0021$   & $0.4275 \pm 0.0037$   & $0.4149 \pm 0.0073$   & $0.4340 \pm 0.0033$   \\
$\phi (2)$ (deg)    & $10.79 \pm 0.54$      & $8.87 \pm 0.69$       & $7.65 \pm 0.97$       & $9.50 \pm 0.62$       \\
$\mathrm{PA}(2)$ (deg) & $30.12 \pm 0.99$   & $31.08 \pm 1.11$      & $33.62 \pm 2.30$      & $29.91 \pm 1.23$      \\
\enddata
\end{deluxetable*}

\begin{deluxetable*}{lrrrr}
\caption{Best-fit Parameters of the Light-curve Model $(u_0<0)$ for Individual Data Sets\label{tab:phot_params_all2}}
\tablehead{\colhead{Parameter} & \colhead{OGLE} & \colhead{KMTC} & \colhead{KMTS} & \colhead{KMTA}}
\startdata
$t_0$ (HJD')        & $179.8150 \pm 0.0082$ & $179.8063 \pm 0.0304$ & $179.8206 \pm 0.0393$ & $179.8324 \pm 0.0193$ \\
$u_0$               & $-0.01098 \pm 0.00010$ & $-0.00798 \pm 0.00434$ & $-0.00851 \pm 0.00461$ & $-0.00799 \pm 0.00113$ \\
$\tE$ (days)           & $93.29 \pm 0.53$      & $91.10 \pm 0.73$      & $93.68 \pm 1.45$      & $93.75 \pm 0.78$      \\
$\piEN$             & $0.3699 \pm 0.0198$   & $0.3172 \pm 0.0176$   & $0.2557 \pm 0.0299$   & $0.3281 \pm 0.0219$   \\
$\piEE$             & $0.1243 \pm 0.0039$   & $0.1252 \pm 0.0046$   & $0.1190 \pm 0.0088$   & $0.1183 \pm 0.0050$   \\
\hline
$\chi^2$/dof     & $541.9/541$           & $600.7/632$           & $329.3/329$           & $416.9/417$           \\
\hline
$\piE$              & $0.3902 \pm 0.0178$   & $0.3410 \pm 0.0165$   & $0.2822 \pm 0.0277$   & $0.3489 \pm 0.0201$   \\
$\Phi_{\pi}$ (deg)        & $18.56 \pm 1.43$      & $21.55 \pm 1.32$      & $24.98 \pm 2.93$      & $19.84 \pm 1.71$      \\
$u(1)$              & $0.2587 \pm 0.0014$   & $0.2655 \pm 0.0021$   & $0.2588 \pm 0.0039$   & $0.2584 \pm 0.0020$   \\
$\eta (1)$       & $0.5969 \pm 0.0017$   & $0.5890 \pm 0.0024$   & $0.5969 \pm 0.0045$   & $0.5973 \pm 0.0024$   \\
$\phi (1)$ (deg)    & $176.93 \pm 0.32$     & $177.18 \pm 0.89$     & $178.12 \pm 0.93$     & $176.89 \pm 0.38$     \\
$\mathrm{PA}(1)$ (deg) & $-158.36 \pm 1.12$ & $-155.60 \pm 1.55$    & $-153.11 \pm 2.85$    & $-157.05 \pm 1.47$    \\
$u(2)$              & $0.4235 \pm 0.0025$   & $0.4319 \pm 0.0038$   & $0.4178 \pm 0.0076$   & $0.4194 \pm 0.0040$   \\
$\eta (2)$       & $0.4314 \pm 0.0021$   & $0.4243 \pm 0.0032$   & $0.4362 \pm 0.0064$   & $0.4349 \pm 0.0034$   \\
$\phi (2)$ (deg)    & $8.20 \pm 0.54$       & $7.03 \pm 0.57$       & $5.54 \pm 0.80$       & $7.58 \pm 0.57$       \\
$\mathrm{PA}(2)$ (deg) & $10.36 \pm 1.96$   & $14.51 \pm 1.65$      & $19.44 \pm 3.48$      & $12.25 \pm 2.24$      \\
\enddata
\end{deluxetable*}

The light-curve model of a microlensing event makes it possible to predict the brightness ratio $\eta$ between the minor and major image at a given epoch of interferometric observations. Moreover, the detection of the microlensing parallax in the light curve allows us to predict the orientation of microlensed images in the sky. The position angle (PA; north through east) of the microlensing parallax vector can be calculated using the formula 
\begin{equation}
\Phi_{\pi} = \arctan\frac{\piEE}{\piEN}.
\end{equation}
The angle $\phi$ between the source--lens relative proper motion and source--lens relative position at a given time can also be calculated directly from the parameters of the light-curve model:
\begin{equation}
\phi (t) = \arctan\frac{\beta(t)}{\tau(t)}.
\end{equation}
Note that in the limit of no parallax, this equation simplifies to $\phi(t)=\arctan(u_0\tE/(t-t_0))$, which is equivalent to Equation~(8) derived by \citet{dong2019}. Then, following \citet{dong2019}, the position angle of the minor image relative to the major image (north through east) is simply $\mathrm{PA} = \Phi_{\pi} + \phi$ if $u_0>0$ or $\mathrm{PA} = \Phi_{\pi} - \phi$ if $u_0<0$. These two cases depend on the trajectory of the lens relative to the source: $u_0>0$ if the lens passes the source on its right and $u_0<0$ on its left \citep{skowron2011}. Note that \citet{dong2019} introduced a different definition of the position angle of the major image relative to the minor image: $\psi=\mathrm{PA}+\pi$.

Table~\ref{tab:phot_params} presents the predicted lens--source separation $u$ (in Einstein radius units), the flux ratio between the minor and major image $\eta$, and the position angle of the images $\mathrm{PA}$ for both epochs of VLTI observations. Note that the expected lens--source separation and flux ratio of the images are virtually identical for the $u_0>0$ and \mbox{$u_0<0$} models. However, the expected position angles are different. For the positive $u_0$ solution, we can predict $\mathrm{PA}=-164.83 \pm 0.95$\,deg and $\mathrm{PA}=31.52 \pm 0.58$\,deg for both VLTI epochs. For the negative $u_0$ solution, we predict $\mathrm{PA}=-156.60 \pm 0.68$\,deg and $\mathrm{PA}=14.29 \pm 1.06$\,deg, respectively.

We also note that the flux ratio of the images is tightly constrained by the light-curve model, with a precision better than 0.3\%. Conversely, the uncertainty of the position angle is dominated by the uncertainty in the determination of the angle $\Phi_{\pi}$, which itself is dominated by the uncertainty of $\piEN$. As we observed small systematic differences between $\piEN$ determined using different data sets (Figure~\ref{fig:parallaxes}), these differences propagate to systematic variations in $\Phi_{\pi}$ and therefore $\mathrm{PA}$. In particular, the angle $\Phi_{\pi}$ can vary from $18.56$ to $25.90$\,deg (Tables~\ref{tab:phot_params_all} and~\ref{tab:phot_params_all2}).

\section{Closure-phase Models}

The VLTI/GRAVITY observations of OGLE-2023-BLG-0061/KMT-2023-BLG-0496 provide a complementary view of the microlensing event. In particular, the spatial resolution of the interferometer allows us to resolve the microlensed images, providing a precise measurement of the angular Einstein radius. This information is not included in the light curve of the event.
We first separately analyze the VLTI data collected during epochs 1a, 1b, and 2, because they were taken at different times and using different instrument configurations. That will allow us to study the consistency between the model parameters derived using different data sets and the consistency with the light-curve model (Sections~\ref{sec:binary_star_model}, \ref{sec:luminous_lens_model}, and \ref{sec:luminous_blend_model}). In Section~\ref{sec:combined_models}, we combine all VLTI data sets to derive the final parameters of the system.
The results presented in this section were independently checked using the PMOIRED software\footnote{\url{https://github.com/amerand/PMOIRED}} \citep{merand2022} and we found virtually identical results.

\subsection{No-lens-light Model}
\label{sec:binary_star_model}

In the simplest case, if the lens is dark and there are no other blended stars in the GRAVITY field of view, the two images of the source created by microlensing look like a mundane ``binary star.'' We start by fitting such a simple binary star model to the closure-phase data. We assume that both images of the source star can be considered as point-like. We place the major image in the origin of the coordinate system, whereas the position of the minor image is parameterized by a vector $(\Delta\alpha,\Delta\delta)$. If we denote the flux ratio between the minor and major image as $\eta$, then the complex visibility is
\begin{equation}
\hat{V} = \frac{1+\eta\exp\left(-\frac{2\pi i}{\lambda}\left(U\Delta\alpha + V \Delta\delta \right)\right)}{1+\eta},
\end{equation}
where $\lambda$ is the wavelength of observations and $(U,V)$ is the separation of the telescopes in the UV plane. The visibility can be calculated for all possible pairs of telescopes ($\hat{V}_{1,2}$,$\hat{V}_{1,3}$,$\hat{V}_{1,4}$,\dots, where the indices denote different telescopes). For a triangle of telescopes, one can define the bispectrum as the product $B_{1,2,3}=\hat{V}_{1,2}\hat{V}_{2,3}\hat{V}_{3,1}$, and then the closure phase (denoted hereafter as T3) is the argument of this bispectrum $\textrm{T3}_{1,2,3}=\arg{(B_{1,2,3})}$ \citep[see,][]{dalal2003}. Closure phases can be calculated for all possible triangles formed by the telescopes of the interferometer. For VLTI, there are four possible closure phases. In theory, one of them is not independent. However, because the measured closure phases may be affected by noise in the data, we employ all four closure-phase sets in the fits.

We first conduct the grid search to find the initial parameters of the model. We keep the flux ratio fixed to $\eta=0.591$ (epoch~1) and $\eta=0.426$ (epoch~2)---that is, the best-fit values from the light-curve model (Table~\ref{tab:phot_params}). We then evaluate the goodness-of-the-fit statistic $\chi^2$ on a grid of $201\times 201$ points uniformly distributed over the range $-10\,\textrm{mas}\leq (\Delta\alpha,\Delta\delta) \leq 10$\,mas. For each case, we find there is only one significant minimum of the $\chi^2$ surface.

We then refine the results of the grid search by allowing all parameters to vary. This is achieved by maximizing the log-likelihood function defined as
\begin{equation}
\begin{split}
\ln\mathcal{L} =& -\frac{1}{2}\sum_{i=1}^{4n_{\rm exp}}\sum_{j=1}^{\Lambda}\frac{(\mathrm{T3}_{ij}-\mathrm{T3}_{ij}^{\rm model})^2}{\sigma(\mathrm{T3})^2_{ij}+\sigma^2_0} + \\
&- \frac{1}{2}\sum_{i=1}^{4n_{\rm exp}}\sum_{j=1}^{\Lambda} \ln\left(\sigma(\mathrm{T3})^2_{ij}+\sigma^2_0\right),
\end{split}
\end{equation}
where $n_{\rm exp}$ is the number of exposures in the epoch and $\Lambda$ is the number of spectral channels. Because the error bars calculated from the pipeline may be underestimated, we add a constant error term $\sigma_0$ in quadrature. The best-fit parameters and their uncertainties are calculated using the Markov Chain Monte Carlo (MCMC) algorithm coded by \citet{foreman2013}. We assume flat (noninformative) priors on all parameters of the model.

\begin{deluxetable}{lrrr}
\caption{Best-fit Parameters of the Closure-phase Model Without the Lens Light \label{tab:cp_params}}
\tablehead{\colhead{Parameter} & \colhead{Epoch 1a} & \colhead{Epoch 1b} & \colhead{Epoch 2}}
\startdata
$\Delta\alpha$ (mas) & $-0.760 \pm 0.012$  & $-0.652 \pm 0.030$     & $1.163 \pm 0.007$\\
$\Delta\delta$ (mas) & $-2.467 \pm 0.018$  & $-2.545 \pm 0.029$     & $2.309 \pm 0.022$\\
$\eta$               & $0.6168 \pm 0.0096$ & $0.6358 \pm 0.0127$    & $0.4414 \pm 0.0098$\\
$\sigma_0$ (deg)     & $7.04 \pm 0.33$     & $2.76^{+0.46}_{-0.40}$ & $2.56 \pm 0.26$\\
\hline
$s$ (mas)            & $2.581 \pm 0.015$   & $2.627 \pm 0.022$      & $2.585 \pm 0.020$\\
$\mathrm{PA}$ (deg)  & $-162.87 \pm 0.36$  & $-165.63 \pm 0.77$     & $26.74 \pm 0.27$\\
$u$                  & $0.2424 \pm 0.0075$ & $0.2266 \pm 0.0107$    & $0.4120 \pm 0.0119$\\
$\thetaE$ (mas)      & $1.2814 \pm 0.0066$ & $1.3050 \pm 0.0107$    & $1.2660 \pm 0.0083$ \\
\hline
$\chi^2$/dof      & $824.5/924$         & $49.6/48$              & $282.6/308$\\
\enddata
\end{deluxetable}

\begin{figure*}[htb]
\includegraphics[width=\textwidth]{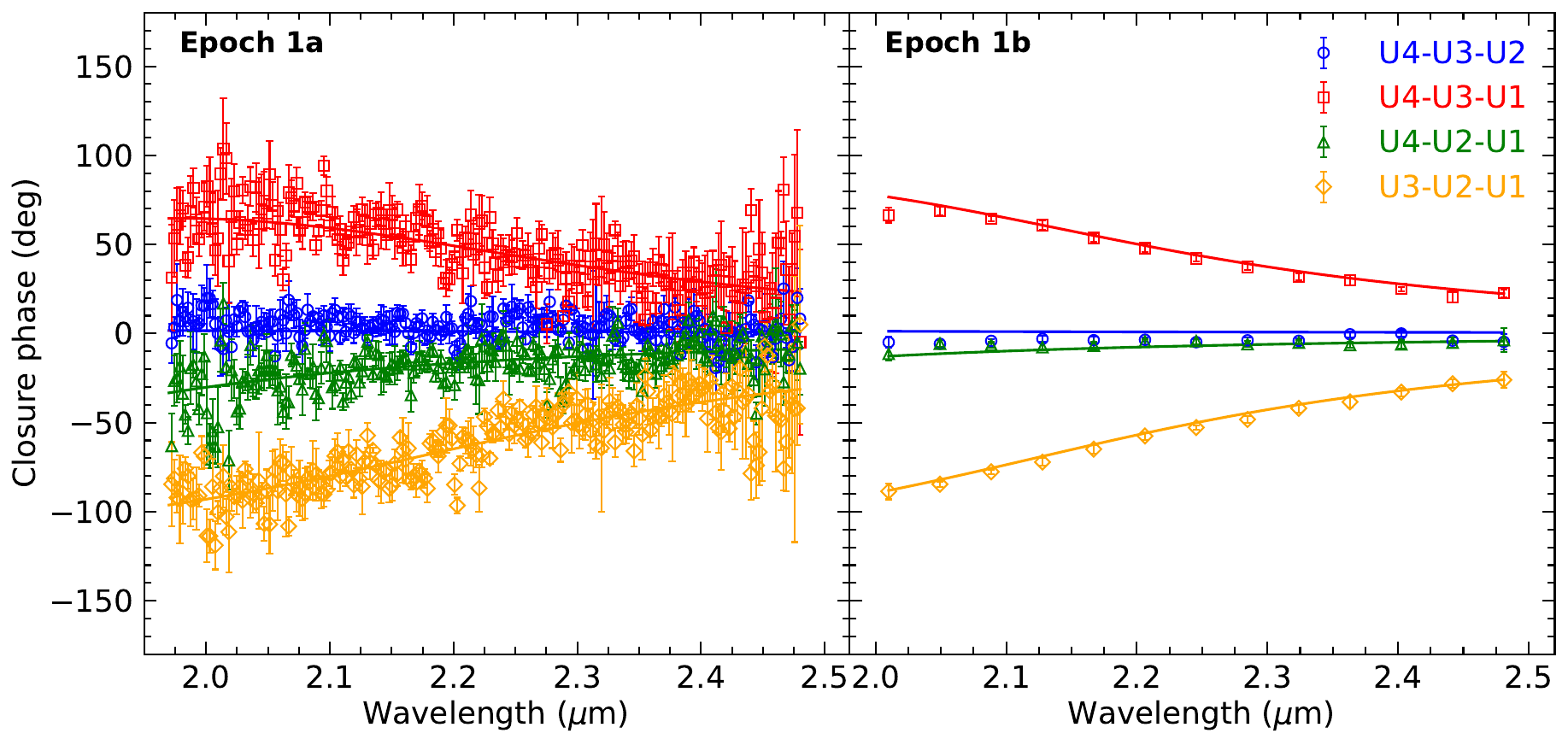}
\caption{Closure-phase data for the VLTI epoch 1. The colored solid lines mark the best-fit microlensing model without the lens light.}
\label{fig:cp1}
\end{figure*}

\begin{figure*}[htb]
\includegraphics[width=\textwidth]{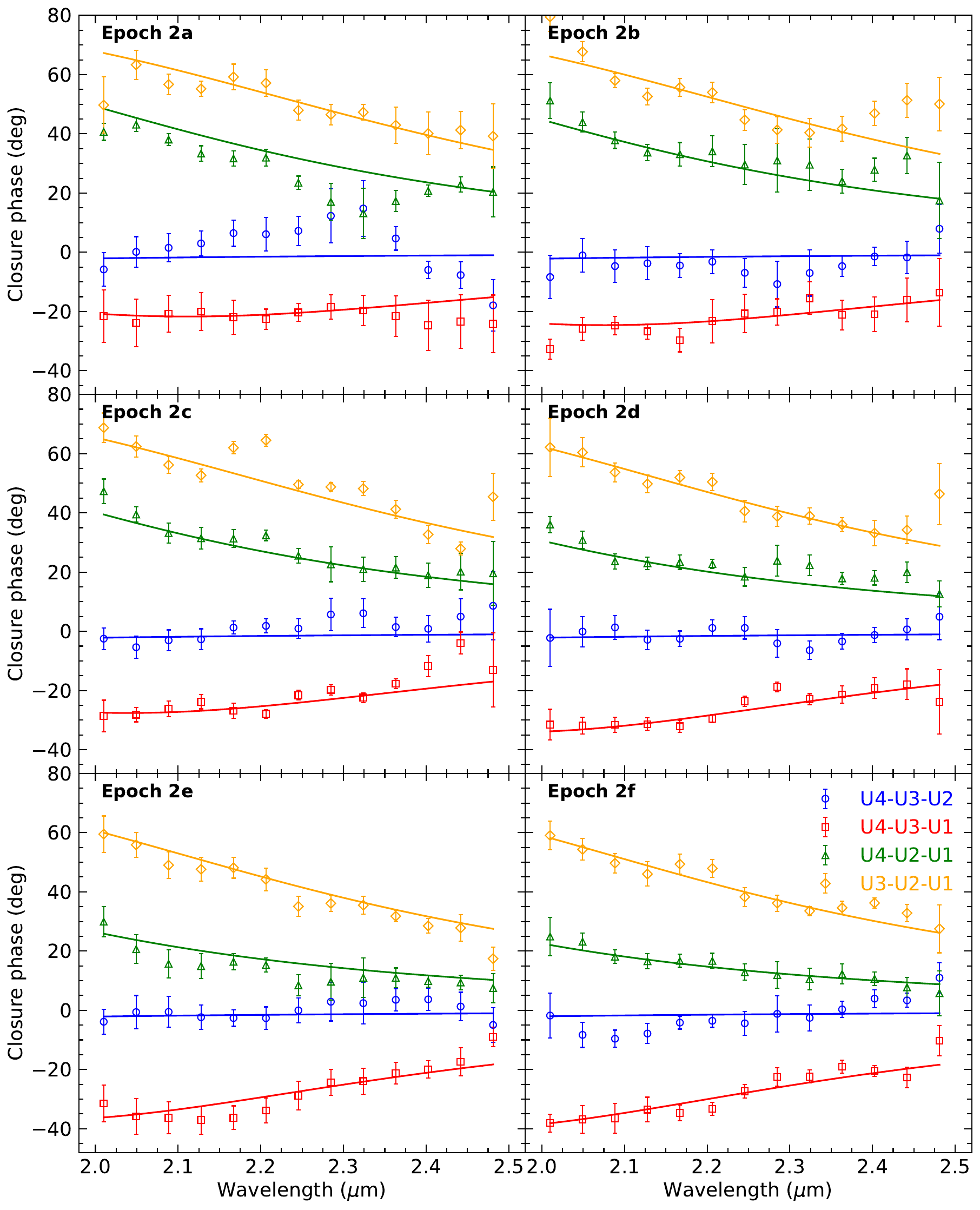}
\caption{Closure-phase data for the VLTI epoch 2. The colored solid lines mark the best-fit microlensing model without the lens light.}
\label{fig:cp2}
\end{figure*}

The results of the fits---separately for VLTI epochs 1a, 1b, and 2---are reported in Table~\ref{tab:cp_params}. The best-fitting closure-phase models are shown in Figures~\ref{fig:cp1} and~\ref{fig:cp2}. In addition to four model parameters, we report four derived quantities: the separation between the images $s=\sqrt{\Delta\alpha^2+\Delta\delta^2}$, their position angle (north through east) $\mathrm{PA}=\arctan \Delta\alpha/\Delta\delta$, the source--lens separation $u$ (in Einstein radius units), and the angular Einstein radius $\thetaE$. The latter quantity is calculated using the formula
\begin{equation}
\thetaE = \frac{s \eta^{1/4}}{\sqrt{\eta}+1}
\end{equation}
derived under the assumption that the flux ratio between the microlensed images determines their separation in Einstein radius units. Similarly, the source--lens separation is
\begin{equation}
u = \frac{1-\sqrt{\eta}}{\eta^{1/4}}.
\end{equation}
The results of the fits to the individual epoch~2 exposures are presented in Table~\ref{tab:cp_params_v2}. 

There is some tension between the medium- and low-resolution interferograms obtained during epoch 1: the measured $\Delta\alpha$ position differs by $3.3\sigma$, the $\Delta\delta$ position by $2.3\sigma$, and the position angle by$3.2\sigma$. On the other hand, the inferred angular Einstein radii are formally consistent between the three different data sets, although the differences between individual measurements can amount up to $1.5\sigma\mbox{--}2.9\sigma$. These tensions indicate that at least one of the analyzed data sets may suffer from unaccounted-for, low-level systematic errors. Another explanation involves possible correlated noise in the closure-phase data that was not taken into account in evaluating the likelihood function \citep[see][]{kammerer2020}. Thus, the error bars reported in Tables~\ref{tab:cp_params} and~\ref{tab:cp_params_v2} may be underestimated.

Further checks are possible because the flux ratio $\eta$ and position angle $\mathrm{PA}$ of the images can be independently measured using the light-curve data (Section~\ref{sec:light_curve_model}). In particular, there are two possible light-curve models differing by the sign of $u_0$, which predict the position angles of the microlensed images as $(-164.8^{\circ}, 31.5^{\circ})$ (positive $u_0$) or $(-156.6^{\circ}, 14.3^{\circ})$ (negative $u_0$), during the VLTI epochs 1 and 2, respectively. The measured angles ($-162.9^{\circ}, 26.7^{\circ}$) (Table~\ref{tab:cp_params}) seem to favor the $u_0>0$ model.

\begin{figure*}[htb]
\centering
\includegraphics[width=\textwidth]{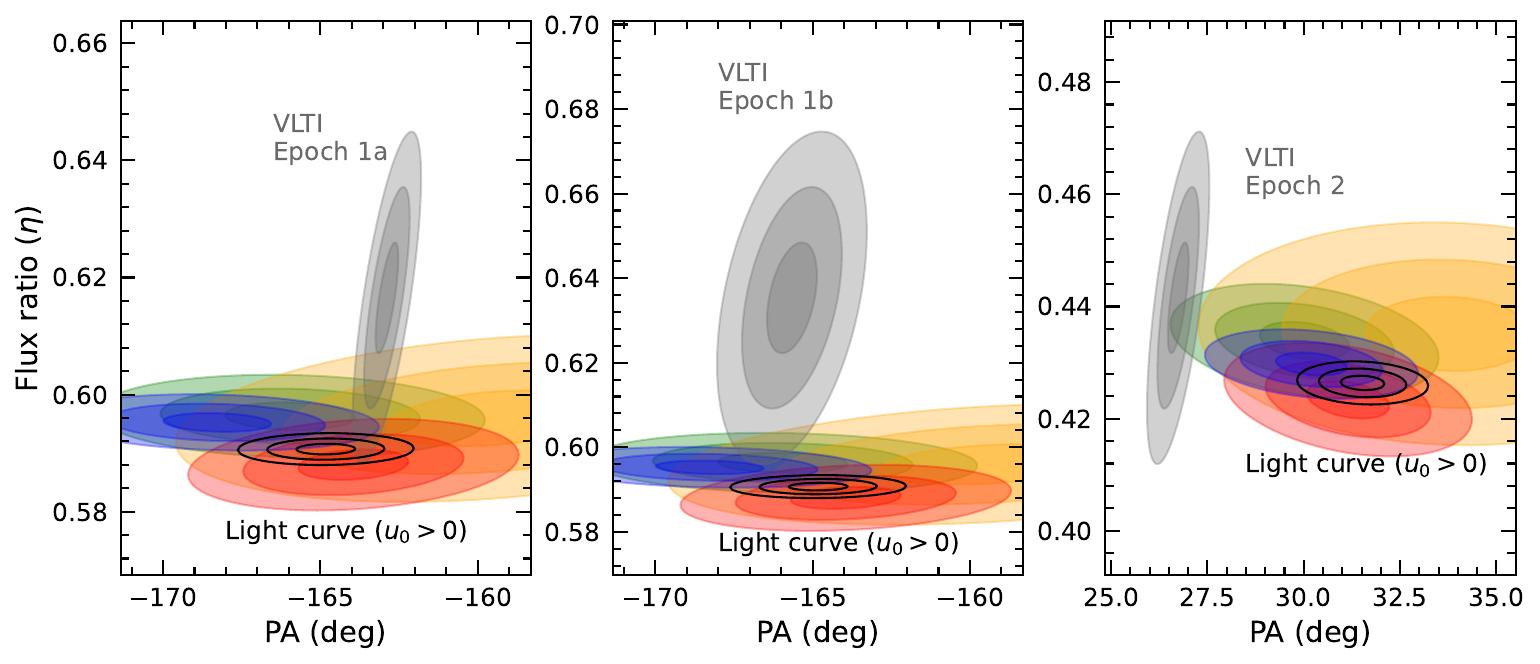}
\caption{Constraints on the flux ratio and position angle of the microlensed images measured from the VLTI data (gray contours). The colored contours mark the constraints from the individual photometric data sets (blue---OGLE; red---KMTC; orange---KMTS; and green---KMTA), while the black contours are measured using all photometric data combined.}
\label{fig:angle_flux_ratio_correlation}
\end{figure*}

However, while the light-curve and closure-phase models agree well for epoch 1, the expected position angles differ by almost $4.8^{\circ}$ (that is, $7.5\sigma$) during epoch 2. That is illustrated in Figure~\ref{fig:angle_flux_ratio_correlation}, which presents $1\sigma$, $2\sigma$, and $3\sigma$ confidence ellipses in the $(\mathrm{PA},\eta)$ plane. The gray contours mark the constraints from three different VLTI data sets, whereas the black contours are calculated based on the light-curve model ($u_0>0$). The blue, red, green, and orange contours mark the constraints on $(\mathrm{PA},\eta)$ from OGLE, KMTC, KMTA, and KMTS data, respectively. 

The position angle of the images measured from interferometric observations can be projected onto the $(\piEN,\piEE)$ plane (Figure~\ref{fig:parallaxes}). The dotted and dashed lines in this figure correspond to the best-fitting values of the PAs during epochs 1a and 2, respectively. The dashed line does not intersect color contours from the light-curve model, which further exemplifies the tension discussed above.

The light-curve model also predicts the flux ratios between the microlensed images: $\eta=0.5907 \pm 0009$ for epoch 1 and $\eta=0.4264 \pm 0.0013$ for epoch 2. The former value is in $2.7\sigma\mbox{--}3.5\sigma$ tension with the closure-phase fits to epoch 1a and 1b data, respectively.

\begin{deluxetable*}{lrrrrrr}
\caption{Best-fit Parameters of the Closure-phase Model for Individual Epoch~2 Data Without the Lens Light\label{tab:cp_params_v2}}
\tablehead{\colhead{Parameter} & \colhead{Epoch 2a} & \colhead{Epoch 2b} & \colhead{Epoch 2c} & \colhead{Epoch 2d} & \colhead{Epoch 2e} & \colhead{Epoch 2f}}
\startdata
$\Delta\alpha$ (mas) & $1.114 \pm 0.016$  & $1.165 \pm 0.017$     & $1.200 \pm 0.015$ & $1.200 \pm 0.012$  & $1.115 \pm 0.022$     & $1.179 \pm 0.018$\\
$\Delta\delta$ (mas) & $2.429 \pm 0.085$  & $2.374 \pm 0.082$     & $2.252 \pm 0.054$ & $2.283 \pm 0.054$  & $2.248 \pm 0.045$     & $2.335 \pm 0.054$\\
$\eta$               & $0.4010 \pm 0.0238$ & $0.4300 \pm 0.0253$    & $0.4698 \pm 0.0253$ & $0.4440 \pm 0.0248$ & $0.3649 \pm 0.0306$    & $0.4289 \pm 0.0244$\\
$\sigma_0$ (deg)     & $1.35 ^{+0.95}_{-0.87}$  & $2.23^{+1.07}_{-1.15}$ & $3.53 ^{+0.65}_{-0.56}$ & $1.34 ^{+0.58}_{-0.62}$  & $0.56^{+0.59}_{-0.39}$ & $1.71 ^{+0.65}_{-0.67}$\\
\hline
$s$ (mas)            & $2.672 \pm 0.079$   & $2.644 \pm 0.075$      & $2.552 \pm 0.047$ & $2.579 \pm 0.048$   & $2.509 \pm 0.035$      & $2.615 \pm 0.047$\\
$u$                  & $0.4608 \pm 0.0295$ & $0.4252 \pm 0.0292$    & $0.3800 \pm 0.0283$ & $0.4088 \pm 0.0269$ & $0.3649 \pm 0.0306$    & $0.4289 \pm 0.0244$\\
$\mathrm{PA}$ (deg)  & $24.64 \pm 0.77$  & $26.13 \pm 0.80$     & $28.04 \pm 0.65$ & $27.71 \pm 0.60$  & $26.36 \pm 0.82$     & $26.77 \pm 0.70$\\
$\thetaE$ (mas)      & $1.3021 \pm 0.0343$ & $1.2933 \pm 0.0328$    & $1.2535 \pm 0.0201$ & $1.2633 \pm 0.0199$ & $1.2342 \pm 0.0142$    & $1.2790 \pm 0.0193$ \\
\hline
$\chi^2$/dof      & $51.6/48$         & $50.6/48$              & $47.5/48$ & $40.2/48$         & $28.6/48$              & $46.5/48$\\
\enddata
\end{deluxetable*}

\subsection{Luminous Lens Model}
\label{sec:luminous_lens_model}

Despite the slight tensions discussed above, the simple binary star model overall describes the closure-phase data well (Figures~\ref{fig:cp1} and~\ref{fig:cp2}). We now modify the model to include the possibility of blended light coming from the lens itself or a luminous companion (either to the lens or source). We change the primary parameters of the model to the lens--source separation $u$, the position angle of the lens relative to the source $\mathrm{PA}$ (which is equal to the position angle of the minor image relative to the major image), and the angular Einstein radius $\thetaE$, which provide a more natural description of a microlensing event than the binary star model parameters. We keep $\sigma_0$ fixed at 6.990, 2.593, and 2.509\,deg for epochs 1a, 1b, and 2, respectively.

We first consider models with a luminous lens. We place the lens in the origin of the coordinate system and calculate the positions of the minor and major image relative to it \citep{dong2019}. We parameterize the ratio of the lens flux to the (unmagnified) source flux as $\eta_{\rm b}$.

We find that the closure-phase data do not provide strong evidence for the light from the lens. Including the lens light in the fits improves the $\chi^2$ by 1.3, 0.1, and 0.9 for epochs 1a, 1b, and 2, respectively. The corresponding 95\% upper limits on $\eta_{\rm b}$ are 0.74, 0.81, and 0.44, respectively. Moreover, we notice that the lens flux is correlated with the projected lens--source separation (and so the flux ratio between the microlensed images) and the angular Einstein radius. 

Thus, we explore the possibility that the light from the lens may be a source of the tensions between the photometric and interferometric data discussed above. We repeat the modeling, taking into account the priors on $u$ from the light-curve model. The best-fit parameters are reported in Table~\ref{tab:cp_params2}. While the tension between $u$ determined from the light-curve model and that from the closure-phase model is removed, the position angles are still slightly different.

\begin{deluxetable}{lrrr}
\caption{Best-fit Parameters of the Luminous Lens Closure-phase Model\label{tab:cp_params2}}
\tablehead{\colhead{Parameter} & \colhead{Epoch 1a} & \colhead{Epoch 1b} & \colhead{Epoch 2}}
\startdata
$u$                  & $0.2640 \pm 0.0008$ & $0.2640 \pm 0.0008$    & $0.4294 \pm 0.0016$\\
$\mathrm{PA}$ (deg)  & $-162.88 \pm 0.36$  & $-165.67 \pm 0.72$     & $26.77 \pm 0.26$\\
$\thetaE$ (mas)      & $1.3266 \pm 0.0114$ & $1.3844 \pm 0.0171$    & $1.2868 \pm 0.0066$ \\
$\eta_{\rm b}$       & $0.215 \pm 0.078$ & $0.370 \pm 0.102$        & $0.0662 \pm 0.0394$\\
\hline
$\eta$               & $0.5907 \pm 0.0009$ & $0.5907 \pm 0.0009$    & $0.4264 \pm 0.0013$\\
$\chi^2$/dof      & $824.4/924$         & $49.6/48$              & $282.6/308$\\
\enddata
\end{deluxetable}

\subsection{Luminous Blend Model}
\label{sec:luminous_blend_model}

We also consider the model with a luminous blend, which has two additional parameters compared to the luminous lens model, the offset of the blend $(\Delta\alpha_{\rm b},\Delta\delta_{\rm b})$ relative to the lens in the sky. We searched for possible blends on a grid of $201 \times 201$ positions uniformly spread over the range $-20 \leq \Delta\alpha_{\rm b},\Delta\delta_{\rm b} \leq 20\,\mathrm{mas}$. We kept the position of the blend fixed but allowed the other parameters ($u$, PA, $\thetaE$, and $\eta_{\rm b}$) to vary. The best-fit parameters were found using a downhill approach using the Nelder--Nead algorithm. We adopted a prior on the lens--source separation $u$ from the light-curve model, and we required $\eta_{\rm b} \geq 0$. 

\begin{figure}
\centering
\includegraphics[width=.5\textwidth]{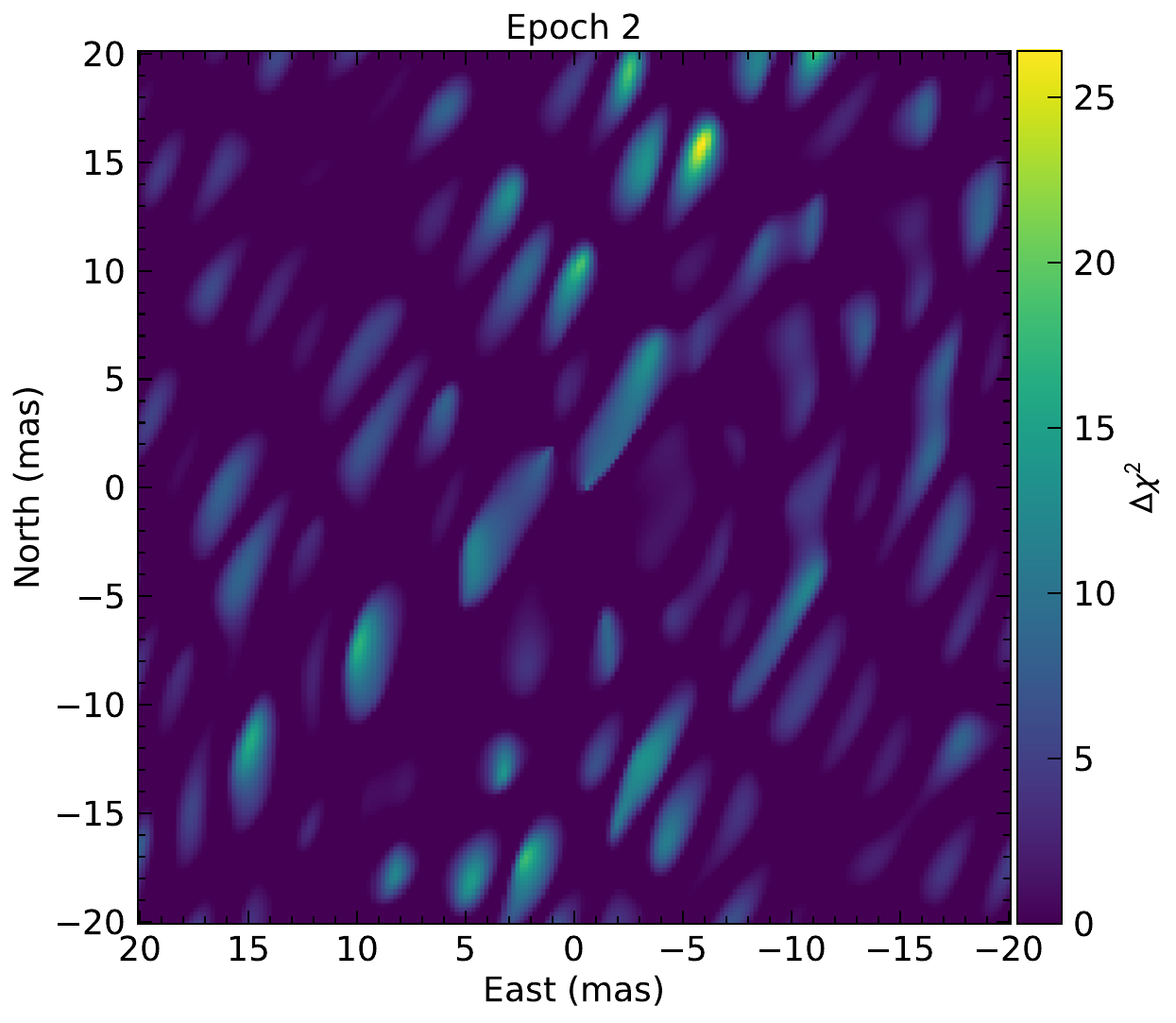}
\caption{Detection map for possible luminous blends in the VLTI epoch 2 data. The color codes the $\chi^2$ improvement over the model without the blend. The lens is located at the origin of the coordinate system.}
\label{fig:grid}
\end{figure}

The grid search results for the epoch~2 data are presented in Figure~\ref{fig:grid}. There is only one local minimum around $(\Delta\alpha_{\rm b},\Delta\delta_{\rm b}) = (-6.0,15.8)\,\mathrm{mas}$ that may be statistically significant (the $\chi^2$ statistics is improved by $\Delta\chi^2=26.6$). We explored this local minimum using the MCMC approach and found $u=0.4295 \pm 0.0016$, $\mathrm{PA}=26.00 \pm 0.24$\,deg, $\thetaE=1.2672 \pm 0.0054$\,mas, $\Delta\alpha_{\rm b}=-5.91 \pm 0.18$\,mas, $\Delta\delta_{\rm b}=15.73 \pm 0.32$\,mas, and $\eta_{\rm b}=0.0426 \pm 0.0094$. Thus, even if we consider this solution statistically significant, the blend does not explain the tension between the position angle measured using the light-curve and interferometric data.

\subsection{Final Closure-phase Models}
\label{sec:combined_models}

Finally, we measure the angular Einstein radius of the event $\thetaE$ using all VLTI data combined. The baseline model has five parameters: the source--lens separations $u$ and position angles of the images PA during epochs 1 and 2 and $\thetaE$ (the same during both epochs). To estimate the impact of systematic errors on the parameter uncertainties, we employ the bootstrapping method \citep[e.g.,][]{efron1979,kervella2004_vinci,lachaume2019}. This involves randomly selecting eight VLTI interferograms (with replacement), feeding them to our fitting routine, and then repeating the procedure multiple times (in this case, 5000 times), to obtain the multivariate probability density function for the parameters of the model. Such a procedure enables us to retrieve more realistic uncertainties on the model parameters. The results of the bootstrapping method are reported in the second column of Table~\ref{tab:cp_params_final}.

We then repeat the modeling, allowing an additional flux from the lens. We consider two models: with and without the prior on $u$ from the light-curve model (the third and fourth columns of Table~\ref{tab:cp_params_final}.) As in Section~\ref{sec:luminous_lens_model}, we find that including the lens light does not significantly improve the fits, and we can only measure upper limits on $\eta_{\rm b}$ (95\% confidence).

\begin{deluxetable}{lrrr}
\caption{Final Closure-phase Models\label{tab:cp_params_final}}
\tablehead{\colhead{Parameter} & \colhead{No Blend} & \colhead{Luminous Lens} & \colhead{Luminous Lens}\\
\colhead{} & \colhead{} & \colhead{} & \colhead{(+ Prior on $u$)}
}
\startdata
$u(1)$                  & $0.2317 \pm 0.0125$ & $0.2550 \pm 0.0217$    & $0.2634 \pm 0.0002$\\
$\mathrm{PA}(1)$ (deg)  & $-163.17 \pm 1.00$  & $-163.68 \pm 1.29$     & $-163.31 \pm 0.81$\\
$u(2)$                  & $0.4286 \pm 0.0122$ & $0.4740 \pm 0.0575$    & $0.4306 \pm 0.0004$\\
$\mathrm{PA}(2)$ (deg)  & $26.40 \pm 0.45$    & $26.79 \pm 0.51$       & $26.60 \pm 0.60$\\
$\thetaE$ (mas)         & $1.2802 \pm 0.0085$ & $1.3239 \pm 0.0351$    & $1.3082 \pm 0.0110$ \\
$\eta_{\rm b}$          & \dots               & $<0.36$      & $<0.23$\\
\enddata
\end{deluxetable}

\section{Physical Parameters of the Lens}

The mass and distance to the lens can be obtained from
\begin{equation}
M = \frac{\thetaE}{\kappa\piE}, \quad D_{\rm l} = \frac{\mathrm{au}}{\piE\thetaE+\mathrm{au}/D_{\rm s}},
\end{equation}
where $D_{\rm s} \approx 8\,\mathrm{kpc}$ is the source distance. As the interferometric data do not provide strong evidence for the light from the lens, we adopt $\thetaE = 1.2802 \pm 0.0085$\,mas (Table~\ref{tab:cp_params_final}) as our final measurement. Using $\piE = 0.3330 \pm 0.0084$, we find $M=0.472 \pm 0.012\,M_{\odot}$, $\pirel=0.426 \pm 0.011$\,mas,  and $D_{\rm l} = 1.81 \pm 0.05$\,kpc. The lens is, therefore, most likely a main-sequence star located in the nearby Galactic disk. Alternatively, it may be a white dwarf. The mass function of field white dwarfs peaks at approximately $0.6\,M_{\odot}$ and falls steeply toward lower masses \citep[e.g.,][]{tremblay2016,torres2021,cunningham2024}. Because white dwarfs less massive than $0.5\,M_{\odot}$ are much less frequent than main-sequence stars of that mass, we find this possibility unlikely.

If the lens is a main-sequence star, its absolute magnitude is $M_K = 5.76 \pm 0.04$ \citep{pecaut2013}, and its apparent $K$-band magnitude is $17.20 \pm 0.07$, assuming $K$-band extinction of 0.15\,mag toward the lens \citep{nataf2013}. That corresponds to $\eta_{\rm b}\approx 0.03$ in the $K$ band, consistent with the limits on the lens flux from the VLTI data.

\begin{figure}
\includegraphics[width=.5\textwidth]{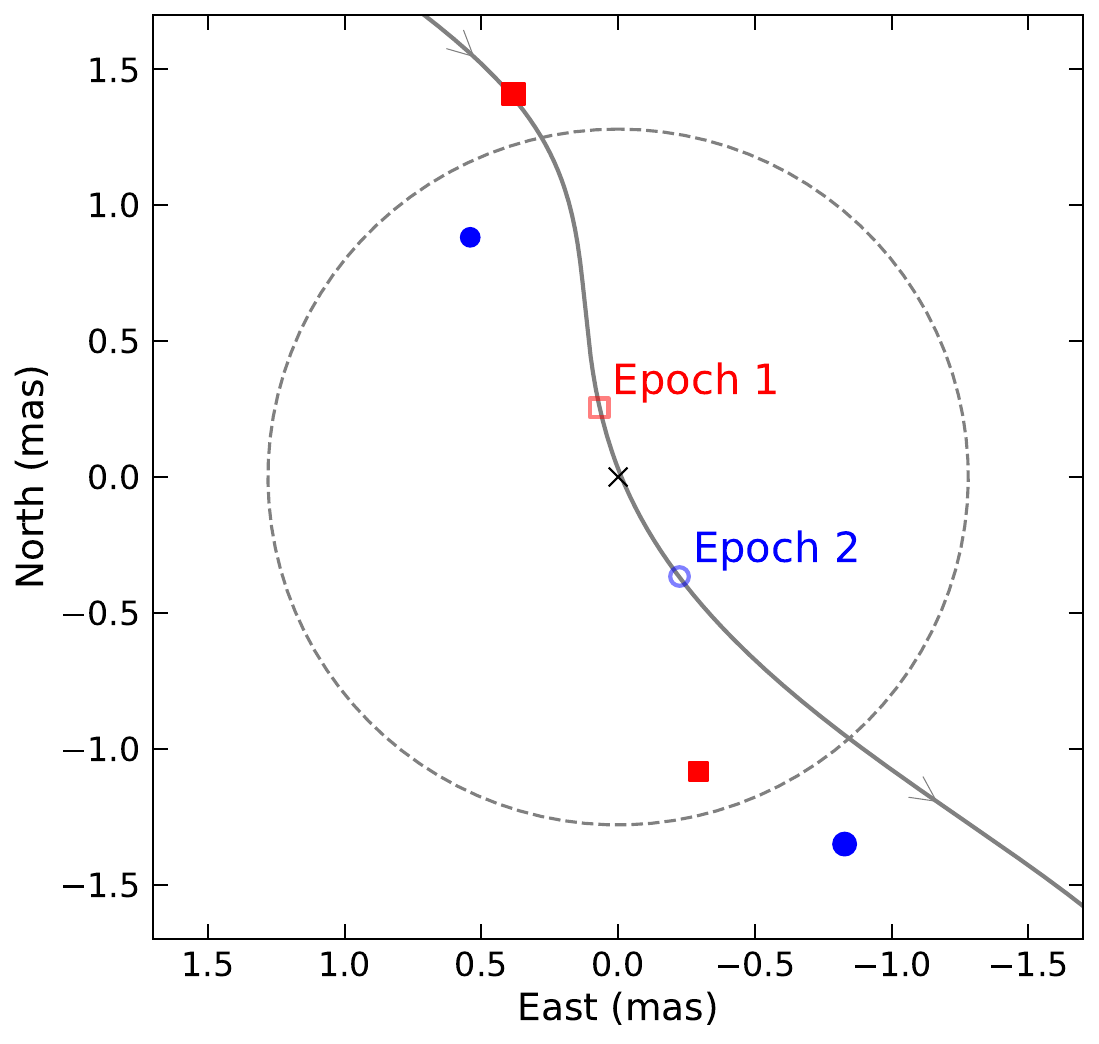}
\caption{Geometry of the event. The gray solid line marks the trajectory of the source relative to the lens (black cross) in the sky. The open symbols (red square and blue circle) mark the positions of the source during the two epochs of VLTI observations. The filled symbols (red squares and blue circles) mark the positions of the two images of the source observed during epoch~1 and epoch~2, respectively. The dashed circle shows the Einstein radius.}
\label{fig:geometry}
\end{figure}

The trajectory of the source in the sky is presented in Figure~\ref{fig:geometry} by a gray solid line. The open symbols (red square and blue circle) mark the positions of the source during the two epochs of VLTI observations, whereas the filled symbols mark the positions of the images of the source.

\section{Discussion and Conclusions}

The start of the scientific operations of GRAVITY Wide opens up completely new possibilities for the follow-up of microlensing events. Only very few of the brightest events could have been observed with the standard on-axis GRAVITY mode. In contrast, GRAVITY Wide observations provide the opportunity for studies of a large sample of events and measurements of masses, distances, and transverse velocities of isolated objects, including neutron stars and black holes. Interferometric observations are very efficient at the same time---only a 1~hr exposure is sufficient for determining the angular Einstein radius. That is in stark contrast to astrometric microlensing measurements, which require time-consuming monitoring lasting several years \citep{sahu2022,lam2022}. Astrometric measurements are also more susceptible to systematic errors due to blending than interferometric observations \citep{mroz2022}. Thus, as argued by \citet{gould2023b}, interferometric observations with GRAVITY Wide (and its successor, GRAVITY+) hold promise for routine detections of isolated neutron stars and stellar remnants.

In this paper, we report the first successful observation of a microlensing event with GRAVITY Wide and the resolution of microlensed images in the event OGLE-2023-BLG-0061/KMT-2023-BLG-0496. These observations serve as a test bed for verifying the capabilities of the new instrument. In particular, some observations were carried out using the low-resolution mode, which had never been tried before with GRAVITY Wide. The comparison of medium- and low-resolution interferograms (epochs 1a and 1b) reveals low-level systematic differences between the inferred parameters of the lensing system (Table~\ref{tab:cp_params}). Similar systematic differences can be seen in models fitted to individual exposures taken during epoch~2 (Table~\ref{tab:cp_params_v2}). Judging from Figures~\ref{fig:cp1} and~\ref{fig:cp2}, the systematic and correlated errors in the closure phase data can reach up to 5--10\,deg. We defer a detailed investigation of the systematic errors in the closure-phase data to a separate study.

We use the bootstrapping method to deal with these systematic and correlated errors. Thanks to the large number of observations, we are still able to measure the angular Einstein radius with a subpercent precision, $\thetaE = 1.2802 \pm 0.0085$\,mas. By combining the information from the light-curve and the closure-phase data, we measure the mass of the lens with a precision of 2.6\%, $M=0.472 \pm 0.012\,M_{\odot}$.

Independent tests of the accuracy of the closure-phase model are possible with the light-curve model. The model of the microlensing event light curve predicts the brightness ratio of the microlensed images and their position angle in the sky. While the predicted flux ratio matches that inferred from the closure-phase data reasonably well (Figure~\ref{fig:angle_flux_ratio_correlation}), there is a tension in the position angles measured using the epoch~2 data. This tension cannot be explained by additional blended light in the closure-phase data (Section~\ref{sec:luminous_blend_model}). 

One possible explanation for this systematic difference is the unaccounted-for systematic errors in the closure-phase data. Alternatively, the problem may lie in the light-curve model. However, we find this unlikely, because the residuals from the best-fit light-curve model are smaller than $\approx 0.01$ mag, and there is no strong evidence that the light-curve model is inadequate. A binary-lens model can be ruled out, because the binary lens would produce additional images of the source, which should have been detected in our grid search. Still, the unaccounted-for orbital motion of the source (the ``xallarap effect'') could, in principle, partly explain the discrepancy. The xallarap effect would not create additional images but would deflect the path of the source on the sky. Our modeling, however, does not provide strong evidence for xallarap in the light-curve data (Appendix \ref{sec:xallarap}).

The additional testing of the model presented in this paper will be possible thanks to the data from the \textit{Gaia} satellite \citep{gaia2016}. OGLE-2023-BLG-0061/KMT-2023-BLG-0496 was observed and alerted by the \textit{Gaia} Photometric Science Alerts system \citep{hodgkin2021} as Gaia23ckg. Time-series astrometric data collected by \textit{Gaia}, when made public later this decade, may reveal deflection due to astrometric microlensing effects and enable independent measurements of the angular Einstein radius.

\section*{Acknowledgements}

We thank David Bennett and \L{}ukasz Wyrzykowski for their comments on the manuscript.

Based on observations collected at the European Southern Observatory under ESO program 108.220D.
This research was funded in part by the National Science Centre, Poland, grants OPUS 2021/41/B/ST9/00252 and SONATA 2023/51/D/ST9/00187 awarded to P.M.
This research is partly supported by the National Natural Science Foundation of China (grant No. 12133005) and the science research grants from the China Manned Space Project with No. CMS-CSST-2021-B12. S.D. acknowledges the New Cornerstone Science Foundation through the XPLORER PRIZE.

GRAVITY+ is developed by the Max Planck Institute for Extraterrestrial Physics, the Institute National des Sciences de l'Univers du CNRS (INSU), with its institutes LESIA / Paris Observatory-PSL, IPAG / Grenoble Observatory, Lagrange / Côte d’Azur Observatory, and CRAL / Lyon Observatory, the Max Planck Institute for Astronomy, the University of Cologne, the CENTRA---Centro de Astrofisica e Gravita\c c\~ao, the University of Southampton, the Katholieke Universiteit Leuven, and the European Southern Observatory.
D.D. has received funding from the European Research Council (ERC) under the European Union's Horizon 2020 research and innovation program (grant agreement CoG---866070).

This research has made use of the KMTNet system operated by the Korea Astronomy and Space Science Institute (KASI) at the three host sites of CTIO in Chile, SAAO in South Africa, and SSO in Australia. Data transfer from the host site to KASI was supported by the Korea Research Environment Open NETwork (KREONET). This research was supported by KASI under the R\&D program (project No. 2024-1-832-01) supervised by the Ministry of Science and ICT. W.Zang, H.Y., S.M., R.K., J.Z., and W.Zhu acknowledge support by the National Natural Science Foundation of China (grant No. 12133005). W.Zang acknowledges the support from the Harvard--Smithsonian Center for Astrophysics through the CfA Fellowship. J.C.Y. and I.-G.S. acknowledge support from U.S. NSF grant No. AST-2108414. Y.S. acknowledges support from BSF grant No. 2020740. The work by C.H. was supported by grants from the National Research Foundation of Korea (2019R1A2C2085965 and 2020R1A4A2002885). J.C.Y. acknowledges support from a Scholarly Studies grant from the Smithsonian Institution.

\bibliographystyle{aasjournal}
\bibliography{pap}

\begin{thebibliography}{}
\expandafter\ifx\csname natexlab\endcsname\relax\def\natexlab#1{#1}\fi

\bibitem[{{Albrow} {et~al.}(2009){Albrow}, {Horne}, {Bramich}, {Fouqu{\'e}},
  {Miller}, {Beaulieu}, {Coutures}, {Menzies}, {Williams}, {Batista},
  {Bennett}, {Brillant}, {Cassan}, {Dieters}, {Dominis Prester}, {Donatowicz},
  {Greenhill}, {Kains}, {Kane}, {Kubas}, {Marquette}, {Pollard}, {Sahu},
  {Tsapras}, {Wambsganss}, \& {Zub}}]{albrow2009}
{Albrow}, M.~D., {Horne}, K., {Bramich}, D.~M., {et~al.} 2009, \mnras, 397,
  2099

\bibitem[{{Arsenault} {et~al.}(2003){Arsenault}, {Alonso}, {Bonnet}, {Brynnel},
  {Delabre}, {Donaldson}, {Dupuy}, {Fedrigo}, {Farinato}, {Hubin}, {Ivanescu},
  {Kasper}, {Paufique}, {Rossi}, {Tordo}, {Stroebele}, {Lizon}, {Gigan},
  {Delplancke}, {Silber}, {Quattri}, \& {Reiss}}]{arsenault2003}
{Arsenault}, R., {Alonso}, J., {Bonnet}, H., {et~al.} 2003, in Society of
  Photo-Optical Instrumentation Engineers (SPIE) Conference Series, Vol. 4839,
  Adaptive Optical System Technologies II, ed. P.~L. {Wizinowich} \&
  D.~{Bonaccini}, 174--185

\bibitem[{{Bachelet} {et~al.}(2022){Bachelet}, {Zieli{\'n}ski}, {Gromadzki},
  {Gezer}, {Rybicki}, {Kruszy{\'n}ska}, {Ihanec}, {Wyrzykowski}, {Street},
  {Tsapras}, {Hundertmark}, {Cassan}, {Harbeck}, \& {Rabus}}]{bachelet2022}
{Bachelet}, E., {Zieli{\'n}ski}, P., {Gromadzki}, M., {et~al.} 2022, \aap, 657,
  A17

\bibitem[{{Cassan} {et~al.}(2022){Cassan}, {Ranc}, {Absil}, {Wyrzykowski},
  {Rybicki}, {Bachelet}, {Le Bouquin}, {Hundertmark}, {Street}, {Surdej},
  {Tsapras}, {Wambsganss}, \& {Wertz}}]{cassan2022}
{Cassan}, A., {Ranc}, C., {Absil}, O., {et~al.} 2022, Nature Astronomy, 6, 121

\bibitem[{{Colavita} {et~al.}(1999){Colavita}, {Wallace}, {Hines}, {Gursel},
  {Malbet}, {Palmer}, {Pan}, {Shao}, {Yu}, {Boden}, {Dumont}, {Gubler},
  {Koresko}, {Kulkarni}, {Lane}, {Mobley}, \& {van Belle}}]{colavita1999}
{Colavita}, M.~M., {Wallace}, J.~K., {Hines}, B.~E., {et~al.} 1999, \apj, 510,
  505

\bibitem[{{Cunningham} {et~al.}(2024){Cunningham}, {Tremblay}, \& {W.
  O'Brien}}]{cunningham2024}
{Cunningham}, T., {Tremblay}, P.-E., \& {W. O'Brien}, M. 2024, \mnras, 527,
  3602

\bibitem[{{Dalal} \& {Lane}(2003)}]{dalal2003}
{Dalal}, N., \& {Lane}, B.~F. 2003, \apj, 589, 199

\bibitem[{{Delplancke}(2008)}]{delplancke2008}
{Delplancke}, F. 2008, \nar, 52, 199

\bibitem[{{Delplancke} {et~al.}(2001){Delplancke}, {G{\'o}rski}, \&
  {Richichi}}]{delplancke2001}
{Delplancke}, F., {G{\'o}rski}, K.~M., \& {Richichi}, A. 2001, \aap, 375, 701

\bibitem[{{Dong} {et~al.}(2019){Dong}, {M{\'e}rand}, {Delplancke-Str{\"o}bele},
  {Gould}, {Chen}, {Post}, {Kochanek}, {Stanek}, {Christie}, {Mutel},
  {Natusch}, {Holoien}, {Prieto}, {Shappee}, \& {Thompson}}]{dong2019}
{Dong}, S., {M{\'e}rand}, A., {Delplancke-Str{\"o}bele}, F., {et~al.} 2019,
  \apj, 871, 70

\bibitem[{Efron(1979)}]{efron1979}
Efron, B. 1979, The Annals of Statistics, 7, 1

\bibitem[{{Einstein}(1936)}]{einstein1936}
{Einstein}, A. 1936, Science, 84, 506

\bibitem[{{Eisenhauer} {et~al.}(2023){Eisenhauer}, {Monnier}, \&
  {Pfuhl}}]{eisenhauer2023}
{Eisenhauer}, F., {Monnier}, J.~D., \& {Pfuhl}, O. 2023, \araa, 61, 237

\bibitem[{{Foreman-Mackey} {et~al.}(2013){Foreman-Mackey}, {Hogg}, {Lang}, \&
  {Goodman}}]{foreman2013}
{Foreman-Mackey}, D., {Hogg}, D.~W., {Lang}, D., \& {Goodman}, J. 2013, \pasp,
  125, 306

\bibitem[{{Gaia Collaboration} {et~al.}(2016){Gaia Collaboration}, {Prusti},
  {de Bruijne}, {Brown}, {Vallenari}, {Babusiaux}, {Bailer-Jones}, {Bastian},
  {Biermann}, {Evans}, \& et~al.}]{gaia2016}
{Gaia Collaboration}, {Prusti}, T., {de Bruijne}, J.~H.~J., {et~al.} 2016,
  \aap, 595, A1

\bibitem[{{Gaia Collaboration} {et~al.}(2023){Gaia Collaboration}, {Vallenari},
  {Brown}, {Prusti}, {de Bruijne}, {Arenou}, {Babusiaux}, {Biermann}, \&
  {Creevey}}]{gaia_dr3}
{Gaia Collaboration}, {Vallenari}, A., {Brown}, A.~G.~A., {et~al.} 2023, \aap,
  674, A1

\bibitem[{{Gould}(1994)}]{gould1994}
{Gould}, A. 1994, \apjl, 421, L71

\bibitem[{{Gould}(2004)}]{gould2004}
{Gould}, A. 2004, \apj, 606, 319

\bibitem[{{Gould}(2023)}]{gould2023b}
{Gould}, A. 2023, arXiv e-prints, arXiv:2310.19164

\bibitem[{{Gould} {et~al.}(1994){Gould}, {Miralda-Escude}, \&
  {Bahcall}}]{gould1994_1dparallax}
{Gould}, A., {Miralda-Escude}, J., \& {Bahcall}, J.~N. 1994, \apjl, 423, L105

\bibitem[{{GRAVITY Collaboration} {et~al.}(2017){GRAVITY Collaboration},
  {Abuter}, {Accardo}, {Amorim}, {Anugu}, {{\'A}vila}, {Azouaoui}, {Benisty},
  {Berger}, {Blind}, {Bonnet}, {Bourget}, {Brandner}, {Brast}, {Buron},
  {Burtscher}, {Cassaing}, {Chapron}, {Choquet}, {Cl{\'e}net}, {Collin},
  {Coud{\'e} Du Foresto}, {de Wit}, {de Zeeuw}, {Deen},
  {Delplancke-Str{\"o}bele}, {Dembet}, {Derie}, {Dexter}, {Duvert}, {Ebert},
  {Eckart}, {Eisenhauer}, {Esselborn}, {F{\'e}dou}, {Finger}, {Garcia}, {Garcia
  Dabo}, {Garcia Lopez}, {Gendron}, {Genzel}, {Gillessen}, {Gonte}, {Gordo},
  {Grould}, {Gr{\"o}zinger}, {Guieu}, {Haguenauer}, {Hans}, {Haubois}, {Haug},
  {Haussmann}, {Henning}, {Hippler}, {Horrobin}, {Huber}, {Hubert}, {Hubin},
  {Hummel}, {Jakob}, {Janssen}, {Jochum}, {Jocou}, {Kaufer}, {Kellner},
  {Kendrew}, {Kern}, {Kervella}, {Kiekebusch}, {Klein}, {Kok}, {Kolb}, {Kulas},
  {Lacour}, {Lapeyr{\`e}re}, {Lazareff}, {Le Bouquin}, {L{\`e}na}, {Lenzen},
  {L{\'e}v{\^e}que}, {Lippa}, {Magnard}, {Mehrgan}, {Mellein}, {M{\'e}rand},
  {Moreno-Ventas}, {Moulin}, {M{\"u}ller}, {M{\"u}ller}, {Neumann}, {Oberti},
  {Ott}, {Pallanca}, {Panduro}, {Pasquini}, {Paumard}, {Percheron}, {Perraut},
  {Perrin}, {Pfl{\"u}ger}, {Pfuhl}, {Phan Duc}, {Plewa}, {Popovic}, {Rabien},
  {Ram{\'\i}rez}, {Ramos}, {Rau}, {Riquelme}, {Rohloff}, {Rousset},
  {Sanchez-Bermudez}, {Scheithauer}, {Sch{\"o}ller}, {Schuhler}, {Spyromilio},
  {Straubmeier}, {Sturm}, {Suarez}, {Tristram}, {Ventura}, {Vincent},
  {Waisberg}, {Wank}, {Weber}, {Wieprecht}, {Wiest}, {Wiezorrek}, {Wittkowski},
  {Woillez}, {Wolff}, {Yazici}, {Ziegler}, \& {Zins}}]{gravity2017}
{GRAVITY Collaboration}, {Abuter}, R., {Accardo}, M., {et~al.} 2017, \aap, 602,
  A94

\bibitem[{{GRAVITY+ Collaboration} {et~al.}(2022){GRAVITY+ Collaboration},
  {Abuter}, {Allouche}, {Amorim}, {Bailet}, {Baub{\"o}ck}, {Berger}, {Berio},
  {Bigioli}, {Boebion}, {Bolzer}, {Bonnet}, {Bourdarot}, {Bourget}, {Brandner},
  {Cl{\'e}net}, {Courtney-Barrer}, {Dallilar}, {Davies}, {Defr{\`e}re},
  {Delboulb{\'e}}, {Delplancke}, {Dembet}, {de Zeeuw}, {Drescher}, {Eckart},
  {{\'E}douard}, {Eisenhauer}, {Fabricius}, {Feuchtgruber}, {Finger},
  {F{\"o}rster Schreiber}, {Garcia}, {Garcia}, {Gao}, {Gendron}, {Genzel},
  {Gil}, {Gillessen}, {Gomes}, {Gont{\'e}}, {Gouvret}, {Guajardo}, {Guieu},
  {Hartl}, {Haubois}, {Hau{\ss}mann}, {Hei{\ss}el}, {Henning}, {Hippler},
  {H{\"o}nig}, {Horrobin}, {Hubin}, {Jacqmart}, {Jochum}, {Jocou}, {Kaufer},
  {Kervella}, {Korhonen}, {Kreidberg}, {Lacour}, {Lagarde}, {Lai},
  {Lapeyr{\`e}re}, {Laugier}, {Le Bouquin}, {Leftley}, {L{\'e}na}, {Lutz},
  {Mang}, {Marcotto}, {Maurel}, {M{\'e}rand}, {Millour}, {More}, {Nowacki},
  {Nowak}, {Oberti}, {Ott}, {Pallanca}, {Pasquini}, {Paumard}, {Perraut},
  {Perrin}, {Petrov}, {Pfuhl}, {Pourr{\'e}}, {Rabien}, {Rau}, {Robbe-Dubois},
  {Rochat}, {Salman}, {Sch{\"o}ller}, {Schubert}, {Schuhler}, {Shangguan},
  {Shimizu}, {Scheithauer}, {Sevin}, {Soulez}, {Spang}, {Stadler}, {Stadler},
  {Straubmeier}, {Sturm}, {Tacconi}, {Tristram}, {Vincent}, {von Fellenberg},
  {Uysal}, {Widmann}, {Wieprecht}, {Wiezorrek}, {Woillez}, {Yazici}, {Young},
  \& {Zins}}]{gravity2022}
{GRAVITY+ Collaboration}, {Abuter}, R., {Allouche}, F., {et~al.} 2022, \aap,
  665, A75

\bibitem[{{Hodgkin} {et~al.}(2013){Hodgkin}, {Wyrzykowski}, {Blagorodnova}, \&
  {Koposov}}]{hodgkin2013}
{Hodgkin}, S.~T., {Wyrzykowski}, L., {Blagorodnova}, N., \& {Koposov}, S. 2013,
  Philosophical Transactions of the Royal Society of London Series A, 371,
  20120239

\bibitem[{{Hodgkin} {et~al.}(2021){Hodgkin}, {Harrison}, {Breedt}, {Wevers},
  {Rixon}, {Delgado}, {Yoldas}, {Kostrzewa-Rutkowska}, {Wyrzykowski}, {van
  Leeuwen}, {Blagorodnova}, {Campbell}, {Eappachen}, {Fraser}, {Ihanec},
  {Koposov}, {Kruszy{\'n}ska}, {Marton}, {Rybicki}, {Brown}, {Burgess},
  {Busso}, {Cowell}, {De Angeli}, {Diener}, {Evans}, {Gilmore}, {Holland},
  {Jonker}, {van Leeuwen}, {Mignard}, {Osborne}, {Portell}, {Prusti},
  {Richards}, {Riello}, {Seabroke}, {Walton}, {{\'A}brah{\'a}m}, {Altavilla},
  {Baker}, {Bastian}, {O'Brien}, {de Bruijne}, {Butterley}, {Carrasco},
  {Casta{\~n}eda}, {Clark}, {Clementini}, {Copperwheat}, {Cropper},
  {Damljanovic}, {Davidson}, {Davis}, {Dennefeld}, {Dhillon}, {Dolding},
  {Dominik}, {Esquej}, {Eyer}, {Fabricius}, {Fridman}, {Froebrich}, {Garralda},
  {Gomboc}, {Gonz{\'a}lez-Vidal}, {Guerra}, {Hambly}, {Hardy}, {Holl},
  {Hourihane}, {Japelj}, {Kann}, {Kiss}, {Knigge}, {Kolb}, {Komossa},
  {K{\'o}sp{\'a}l}, {Kov{\'a}cs}, {Kun}, {Leto}, {Lewis}, {Littlefair},
  {Mahabal}, {Mundell}, {Nagy}, {Padeletti}, {Palaversa}, {Pigulski},
  {Pretorius}, {van Reeven}, {Ribeiro}, {Roelens}, {Rowell}, {Schartel},
  {Scholz}, {Schwope}, {Sip{\H{o}}cz}, {Smartt}, {Smith}, {Serraller},
  {Steeghs}, {Sullivan}, {Szabados}, {Szegedi-Elek}, {Tisserand}, {Tomasella},
  {van Velzen}, {Whitelock}, {Wilson}, \& {Young}}]{hodgkin2021}
{Hodgkin}, S.~T., {Harrison}, D.~L., {Breedt}, E., {et~al.} 2021, \aap, 652,
  A76

\bibitem[{{Hog} {et~al.}(1995){Hog}, {Novikov}, \& {Polnarev}}]{hog1995}
{Hog}, E., {Novikov}, I.~D., \& {Polnarev}, A.~G. 1995, \aap, 294, 287

\bibitem[{{Kammerer} {et~al.}(2020){Kammerer}, {M{\'e}rand}, {Ireland}, \&
  {Lacour}}]{kammerer2020}
{Kammerer}, J., {M{\'e}rand}, A., {Ireland}, M.~J., \& {Lacour}, S. 2020, \aap,
  644, A110

\bibitem[{{Kendrew} {et~al.}(2012){Kendrew}, {Hippler}, {Brandner},
  {Cl{\'e}net}, {Deen}, {Gendron}, {Huber}, {Klein}, {Laun}, {Lenzen},
  {Naranjo}, {Neumann}, {Ramos}, {Rohloff}, {Yang}, {Eisenhauer}, {Amorim},
  {Perraut}, {Perrin}, {Straubmeier}, {Fedrigo}, \& {Suarez
  Valles}}]{kendrew2012}
{Kendrew}, S., {Hippler}, S., {Brandner}, W., {et~al.} 2012, in Society of
  Photo-Optical Instrumentation Engineers (SPIE) Conference Series, Vol. 8446,
  Ground-based and Airborne Instrumentation for Astronomy IV, ed. I.~S.
  {McLean}, S.~K. {Ramsay}, \& H.~{Takami}, 84467W

\bibitem[{{Kervella} {et~al.}(2004){Kervella}, {S{\'e}gransan}, \& {Coud{\'e}
  du Foresto}}]{kervella2004_vinci}
{Kervella}, P., {S{\'e}gransan}, D., \& {Coud{\'e} du Foresto}, V. 2004, \aap,
  425, 1161

\bibitem[{{Kim} {et~al.}(2018){Kim}, {Hwang}, {Shvartzvald}, {Yee}, {Albrow},
  {Cha}, {Chung}, {Gould}, {Han}, {Jung}, {Kim}, {Kim}, {Lee}, {Lee}, {Lee},
  {Park}, {Pogge}, {Ryu}, {Shin}, \& {Zang}}]{kim2018_alerts}
{Kim}, H.-W., {Hwang}, K.-H., {Shvartzvald}, Y., {et~al.} 2018, arXiv e-prints,
  arXiv:1806.07545

\bibitem[{{Kim} {et~al.}(2016){Kim}, {Lee}, {Park}, {Kim}, {Cha}, {Lee}, {Han},
  {Chun}, \& {Yuk}}]{kim2016}
{Kim}, S.-L., {Lee}, C.-U., {Park}, B.-G., {et~al.} 2016, \jkas, 49, 37

\bibitem[{{Lachaume} {et~al.}(2019){Lachaume}, {Rabus}, {Jord{\'a}n}, {Brahm},
  {Boyajian}, {von Braun}, \& {Berger}}]{lachaume2019}
{Lachaume}, R., {Rabus}, M., {Jord{\'a}n}, A., {et~al.} 2019, \mnras, 484, 2656

\bibitem[{{Lacour} {et~al.}(2019){Lacour}, {Dembet}, {Abuter}, {F{\'e}dou},
  {Perrin}, {Choquet}, {Pfuhl}, {Eisenhauer}, {Woillez}, {Cassaing},
  {Wieprecht}, {Ott}, {Wiezorrek}, {Tristram}, {Wolff}, {Ram{\'\i}rez},
  {Haubois}, {Perraut}, {Straubmeier}, {Brandner}, \& {Amorim}}]{lacour2019}
{Lacour}, S., {Dembet}, R., {Abuter}, R., {et~al.} 2019, \aap, 624, A99

\bibitem[{{Lam} {et~al.}(2022){Lam}, {Lu}, {Udalski}, {Bond}, {Bennett},
  {Skowron}, {Mroz}, {Poleski}, {Sumi}, {Szymanski}, {Kozlowski},
  {Pietrukowicz}, {Soszynski}, {Ulaczyk}, {Wyrzykowski}, {Miyazaki}, {Suzuki},
  {Koshimoto}, {Rattenbury}, {Hosek}, {Abe}, {Barry}, {Bhattacharya}, {Fukui},
  {Fujii}, {Hirao}, {Itow}, {Kirikawa}, {Kondo}, {Matsubara}, {Matsumoto},
  {Muraki}, {Olmschenk}, {Ranc}, {Okamura}, {Satoh}, {Ishitani Silva}, {Toda},
  {Tristram}, {Vandorou}, {Yama}, {Abrams}, {Agarwal}, {Rose}, \&
  {Terry}}]{lam2022}
{Lam}, C., {Lu}, J.~R., {Udalski}, A., {et~al.} 2022, \apjl, 933, L23

\bibitem[{{Lam} \& {Lu}(2023)}]{lam2023}
{Lam}, C.~Y., \& {Lu}, J.~R. 2023, \apj, 955, 116

\bibitem[{{Lawrence} {et~al.}(2007){Lawrence}, {Warren}, {Almaini}, {Edge},
  {Hambly}, {Jameson}, {Lucas}, {Casali}, {Adamson}, {Dye}, {Emerson},
  {Foucaud}, {Hewett}, {Hirst}, {Hodgkin}, {Irwin}, {Lodieu}, {McMahon},
  {Simpson}, {Smail}, {Mortlock}, \& {Folger}}]{lawrence2007}
{Lawrence}, A., {Warren}, S.~J., {Almaini}, O., {et~al.} 2007, \mnras, 379,
  1599

\bibitem[{{Lucas} {et~al.}(2008){Lucas}, {Hoare}, {Longmore}, {Schr{\"o}der},
  {Davis}, {Adamson}, {Bandyopadhyay}, {de Grijs}, {Smith}, {Gosling},
  {Mitchison}, {G{\'a}sp{\'a}r}, {Coe}, {Tamura}, {Parker}, {Irwin}, {Hambly},
  {Bryant}, {Collins}, {Cross}, {Evans}, {Gonzalez-Solares}, {Hodgkin},
  {Lewis}, {Read}, {Riello}, {Sutorius}, {Lawrence}, {Drew}, {Dye}, \&
  {Thompson}}]{lucas2008}
{Lucas}, P.~W., {Hoare}, M.~G., {Longmore}, A., {et~al.} 2008, \mnras, 391, 136

\bibitem[{{M{\'e}rand}(2022)}]{merand2022}
{M{\'e}rand}, A. 2022, in Society of Photo-Optical Instrumentation Engineers
  (SPIE) Conference Series, Vol. 12183, Optical and Infrared Interferometry and
  Imaging VIII, ed. A.~{M{\'e}rand}, S.~{Sallum}, \& J.~{Sanchez-Bermudez},
  121831N

\bibitem[{{Minniti} {et~al.}(2010){Minniti}, {Lucas}, {Emerson}, {Saito},
  {Hempel}, {Pietrukowicz}, {Ahumada}, {Alonso}, {Alonso-Garcia}, {Arias},
  {Bandyopadhyay}, {Barb{\'a}}, {Barbuy}, {Bedin}, {Bica}, {Borissova},
  {Bronfman}, {Carraro}, {Catelan}, {Clari{\'a}}, {Cross}, {de Grijs},
  {D{\'e}k{\'a}ny}, {Drew}, {Fari{\~n}a}, {Feinstein}, {Fern{\'a}ndez
  Laj{\'u}s}, {Gamen}, {Geisler}, {Gieren}, {Goldman}, {Gonzalez}, {Gunthardt},
  {Gurovich}, {Hambly}, {Irwin}, {Ivanov}, {Jord{\'a}n}, {Kerins}, {Kinemuchi},
  {Kurtev}, {L{\'o}pez-Corredoira}, {Maccarone}, {Masetti}, {Merlo},
  {Messineo}, {Mirabel}, {Monaco}, {Morelli}, {Padilla}, {Palma}, {Parisi},
  {Pignata}, {Rejkuba}, {Roman-Lopes}, {Sale}, {Schreiber}, {Schr{\"o}der},
  {Smith}, {}, {Soto}, {Tamura}, {Tappert}, {Thompson}, {Toledo}, {Zoccali}, \&
  {Pietrzynski}}]{minniti2010}
{Minniti}, D., {Lucas}, P.~W., {Emerson}, J.~P., {et~al.} 2010, \na, 15, 433

\bibitem[{{Miyamoto} \& {Yoshii}(1995)}]{miyamoto1995}
{Miyamoto}, M., \& {Yoshii}, Y. 1995, \aj, 110, 1427

\bibitem[{{Mr{\'o}z} {et~al.}(2022){Mr{\'o}z}, {Udalski}, \&
  {Gould}}]{mroz2022}
{Mr{\'o}z}, P., {Udalski}, A., \& {Gould}, A. 2022, \apjl, 937, L24

\bibitem[{{Nataf} {et~al.}(2013){Nataf}, {Gould}, {Fouqu{\'e}}, {Gonzalez},
  {Johnson}, {Skowron}, {Udalski}, {Szyma{\'n}ski}, {Kubiak},
  {Pietrzy{\'n}ski}, {Soszy{\'n}ski}, {Ulaczyk}, {Wyrzykowski}, \&
  {Poleski}}]{nataf2013}
{Nataf}, D.~M., {Gould}, A., {Fouqu{\'e}}, P., {et~al.} 2013, \apj, 769, 88

\bibitem[{{Nemiroff} \& {Wickramasinghe}(1994)}]{nemi1994}
{Nemiroff}, R.~J., \& {Wickramasinghe}, W.~A.~D.~T. 1994, \apjl, 424, L21

\bibitem[{{Pecaut} \& {Mamajek}(2013)}]{pecaut2013}
{Pecaut}, M.~J., \& {Mamajek}, E.~E. 2013, \apjs, 208, 9

\bibitem[{{Poleski} \& {Yee}(2019)}]{poleski2019}
{Poleski}, R., \& {Yee}, J.~C. 2019, Astronomy and Computing, 26, 35

\bibitem[{{Rybicki} {et~al.}(2022){Rybicki}, {Wyrzykowski}, {Bachelet},
  {Cassan}, {Zieli{\'n}ski}, {Gould}, {Calchi Novati}, {Yee}, {Ryu},
  {Gromadzki}, {Miko{\l}ajczyk}, {Ihanec}, {Kruszy{\'n}ska}, {Hambsch},
  {Zo{\l}a}, {Fossey}, {Awiphan}, {Nakharutai}, {Lewis}, {Olivares E.},
  {Hodgkin}, {Delgado}, {Breedt}, {Harrison}, {van Leeuwen}, {Rixon}, {Wevers},
  {Yoldas}, {Udalski}, {Szyma{\'n}ski}, {Soszy{\'n}ski}, {Pietrukowicz},
  {Koz{\l}owski}, {Skowron}, {Poleski}, {Ulaczyk}, {Mr{\'o}z}, {Iwanek},
  {Wrona}, {Street}, {Tsapras}, {Hundertmark}, {Dominik}, {Beichman}, {Bryden},
  {Carey}, {Gaudi}, {Henderson}, {Shvartzvald}, {Zang}, {Zhu}, {Christie},
  {Green}, {Hennerley}, {McCormick}, {Monard}, {Natusch}, {Pogge}, {Gezer},
  {Gurgul}, {Kaczmarek}, {Konacki}, {Lam}, {Maskoliunas}, {Pakstiene},
  {Ratajczak}, {Stankeviciute}, {Zdanavicius}, \&
  {Zi{\'o}{\l}kowska}}]{rybicki2022}
{Rybicki}, K.~A., {Wyrzykowski}, {\L}., {Bachelet}, E., {et~al.} 2022, \aap,
  657, A18

\bibitem[{{Sahu} {et~al.}(2022){Sahu}, {Anderson}, {Casertano}, {Bond},
  {Udalski}, {Dominik}, {Calamida}, {Bellini}, {Brown}, {Rejkuba}, {Bajaj},
  {Kains}, {Ferguson}, {Fryer}, {Yock}, {Mroz}, {Kozlowski}, {Pietrukowicz},
  {Poleski}, {Skowron}, {Soszynski}, {Szymanski}, {Ulaczyk}, {Wyrzykowski},
  {Barry}, {Bennett}, {Bond}, {Hirao}, {Ishitani Silva}, {Kondo}, {Koshimoto},
  {Ranc}, {Rattenbury}, {Sumi}, {Suzuki}, {Tristram}, {Vandorou}, {Beaulieu},
  {Marquette}, {Cole}, {Fouque}, {Hill}, {Dieters}, {Coutures},
  {Dominis-Prester}, {Bennett}, {Bachelet}, {Menzies}, {Alb-row}, {Pollard},
  {Gould}, {Yee}, {Allen}, {de Almeida}, {Christie}, {Drummond}, {Gal-Yam},
  {Gorbikov}, {Jablonski}, {Lee}, {Maoz}, {Manulis}, {McCormick}, {Natusch},
  {Pogge}, {Shvartzvald}, {Jorgensen}, {Alsubai}, {Andersen}, {Bozza}, {Calchi
  Novati}, {Burgdorf}, {Hinse}, {Hundertmark}, {Husser}, {Kerins},
  {Longa-Pena}, {Mancini}, {Penny}, {Rahvar}, {Ricci}, {Sajadian}, {Skottfelt},
  {Snodgrass}, {Southworth}, {Tregloan-Reed}, {Wambsganss}, {Wertz}, {Tsapras},
  {Street}, {Bramich}, {Horne}, \& {Steele}}]{sahu2022}
{Sahu}, K.~C., {Anderson}, J., {Casertano}, S., {et~al.} 2022, \apj, 933, 83

\bibitem[{{Shao} \& {Colavita}(1992)}]{shao1992}
{Shao}, M., \& {Colavita}, M.~M. 1992, \aap, 262, 353

\bibitem[{{Shappee} {et~al.}(2014){Shappee}, {Prieto}, {Grupe}, {Kochanek},
  {Stanek}, {De Rosa}, {Mathur}, {Zu}, {Peterson}, {Pogge}, {Komossa}, {Im},
  {Jencson}, {Holoien}, {Basu}, {Beacom}, {Szczygie{\l}}, {Brimacombe},
  {Adams}, {Campillay}, {Choi}, {Contreras}, {Dietrich}, {Dubberley},
  {Elphick}, {Foale}, {Giustini}, {Gonzalez}, {Hawkins}, {Howell}, {Hsiao},
  {Koss}, {Leighly}, {Morrell}, {Mudd}, {Mullins}, {Nugent}, {Parrent},
  {Phillips}, {Pojmanski}, {Rosing}, {Ross}, {Sand}, {Terndrup}, {Valenti},
  {Walker}, \& {Yoon}}]{shappee2014}
{Shappee}, B.~J., {Prieto}, J.~L., {Grupe}, D., {et~al.} 2014, \apj, 788, 48

\bibitem[{{Skowron} {et~al.}(2011){Skowron}, {Udalski}, {Gould}, {Dong},
  {Monard}, {Han}, {Nelson}, {McCormick}, {Moorhouse}, {Thornley}, {Maury},
  {Bramich}, {Greenhill}, {Koz{\l}owski}, {Bond}, {Poleski}, {Wyrzykowski},
  {Ulaczyk}, {Kubiak}, {Szyma{\'n}ski}, {Pietrzy{\'n}ski}, {Soszy{\'n}ski},
  {OGLE Collaboration}, {Gaudi}, {Yee}, {Hung}, {Pogge}, {DePoy}, {Lee},
  {Park}, {Allen}, {Mallia}, {Drummond}, {Bolt}, {{$\mu$}FUN Collaboration},
  {Allan}, {Browne}, {Clay}, {Dominik}, {Fraser}, {Horne}, {Kains}, {Mottram},
  {Snodgrass}, {Steele}, {Street}, {Tsapras}, {RoboNet Collaboration}, {Abe},
  {Bennett}, {Botzler}, {Douchin}, {Freeman}, {Fukui}, {Furusawa}, {Hayashi},
  {Hearnshaw}, {Hosaka}, {Itow}, {Kamiya}, {Kilmartin}, {Korpela}, {Lin},
  {Ling}, {Makita}, {Masuda}, {Matsubara}, {Muraki}, {Nagayama}, {Miyake},
  {Nishimoto}, {Ohnishi}, {Perrott}, {Rattenbury}, {Saito}, {Skuljan},
  {Sullivan}, {Sumi}, {Suzuki}, {Sweatman}, {Tristram}, {Wada}, {Yock}, {MOA
  Collaboration}, {Beaulieu}, {Fouqu{\'e}}, {Albrow}, {Batista}, {Brillant},
  {Caldwell}, {Cassan}, {Cole}, {Cook}, {Coutures}, {Dieters}, {Dominis
  Prester}, {Donatowicz}, {Kane}, {Kubas}, {Marquette}, {Martin}, {Menzies},
  {Sahu}, {Wambsganss}, {Williams}, {Zub}, \& {PLANET
  Collaboration}}]{skowron2011}
{Skowron}, J., {Udalski}, A., {Gould}, A., {et~al.} 2011, \apj, 738, 87

\bibitem[{{Skrutskie} {et~al.}(2006){Skrutskie}, {Cutri}, {Stiening},
  {Weinberg}, {Schneider}, {Carpenter}, {Beichman}, {Capps}, {Chester},
  {Elias}, {Huchra}, {Liebert}, {Lonsdale}, {Monet}, {Price}, {Seitzer},
  {Jarrett}, {Kirkpatrick}, {Gizis}, {Howard}, {Evans}, {Fowler}, {Fullmer},
  {Hurt}, {Light}, {Kopan}, {Marsh}, {McCallon}, {Tam}, {Van Dyk}, \&
  {Wheelock}}]{skrutskie2006}
{Skrutskie}, M.~F., {Cutri}, R.~M., {Stiening}, R., {et~al.} 2006, \aj, 131,
  1163

\bibitem[{{Smith} {et~al.}(2003){Smith}, {Mao}, \& {Paczy{\'n}ski}}]{smith2003}
{Smith}, M.~C., {Mao}, S., \& {Paczy{\'n}ski}, B. 2003, \mnras, 339, 925

\bibitem[{{Torres} {et~al.}(2021){Torres}, {Rebassa-Mansergas}, {Camisassa}, \&
  {Raddi}}]{torres2021}
{Torres}, S., {Rebassa-Mansergas}, A., {Camisassa}, M.~E., \& {Raddi}, R. 2021,
  \mnras, 502, 1753

\bibitem[{{Tremblay} {et~al.}(2016){Tremblay}, {Cummings}, {Kalirai},
  {G{\"a}nsicke}, {Gentile-Fusillo}, \& {Raddi}}]{tremblay2016}
{Tremblay}, P.~E., {Cummings}, J., {Kalirai}, J.~S., {et~al.} 2016, \mnras,
  461, 2100

\bibitem[{{Udalski}(2003)}]{udalski2003}
{Udalski}, A. 2003, \actaa, 53, 291

\bibitem[{{Udalski} {et~al.}(2015){Udalski}, {Szyma{\'n}ski}, \&
  {Szyma{\'n}ski}}]{udalski2015}
{Udalski}, A., {Szyma{\'n}ski}, M.~K., \& {Szyma{\'n}ski}, G. 2015, \actaa, 65,
  1

\bibitem[{{Walker}(1995)}]{walker1995}
{Walker}, M.~A. 1995, \apj, 453, 37

\bibitem[{{Witt} \& {Mao}(1994)}]{witt1994}
{Witt}, H.~J., \& {Mao}, S. 1994, \apj, 430, 505

\bibitem[{{Woillez} {et~al.}(2014){Woillez}, {Wizinowich}, {Akeson},
  {Colavita}, {Eisner}, {Millan-Gabet}, {Monnier}, {Pott}, \&
  {Ragland}}]{woillez2014}
{Woillez}, J., {Wizinowich}, P., {Akeson}, R., {et~al.} 2014, \apj, 783, 104

\bibitem[{{Wo{\'z}niak}(2000)}]{wozniak2000}
{Wo{\'z}niak}, P.~R. 2000, \actaa, 50, 421

\bibitem[{{Wu} {et~al.}(2024{\natexlab{a}}){Wu}, {Dong}, {M{\'e}rand},
  {Kochanek}, {Mr{\'o}z}, {Shangguan}, {Christie}, {Tan}, {Bensby},
  {Bland-Hawthorn}, {Buder}, {Eisenhauer}, {Gould}, {Kos}, {Natusch}, {Sharma},
  {Udalski}, {Woillez}, {Buckley}, {Thompson}, {El Dayem}, {Berdeu}, {Berger},
  {Bourdarot}, {Brandner}, {Davies}, {Defr{\`e}re}, {Dougados}, {Drescher},
  {Eckart}, {Fabricius}, {Feuchtgruber}, {F{\"o}rster Schreiber}, {Garcia},
  {Genzel}, {Gillessen}, {Hei{\ss}el}, {H{\"o}nig}, {Houlle}, {Kervella},
  {Kreidberg}, {Lacour}, {Lai}, {Laugier}, {Le Bouquin}, {Leftley}, {Lopez},
  {Lutz}, {Mang}, {Millour}, {Montarg{\`e}s}, {Nowacki}, {Nowak}, {Ott},
  {Paumard}, {Perraut}, {Perrin}, {Petrov}, {Petrucci}, {Pourre}, {Rabien},
  {Ribeiro}, {Robbe-Dubois}, {Sadun Bordoni}, {Santos}, {Sauter}, {Scigliuto},
  {Shimizu}, {Straubmeier}, {Sturm}, {Subroweit}, {Sykes}, {Tacconi},
  {Vincent}, \& {Widmann}}]{wu2024b}
{Wu}, Z., {Dong}, S., {M{\'e}rand}, A., {et~al.} 2024{\natexlab{a}}, \apj, 977,
  229

\bibitem[{{Wu} {et~al.}(2024{\natexlab{b}}){Wu}, {Dong}, {Yi}, {Liu},
  {El-Badry}, {Gould}, {Wyrzykowski}, {Rybicki}, {Bachelet}, {Christie}, {de
  Almeida}, {Monard}, {McCormick}, {Natusch}, {Zieli{\'n}ski}, {Chen}, {Huang},
  {Liu}, {M{\'e}rand}, {Mr{\'o}z}, {Shangguan}, {Udalski}, {Woillez}, {Zhang},
  {Hambsch}, {Miko{\l}ajczyk}, {Gromadzki}, {Ratajczak}, {Kruszy{\'n}ska},
  {Ihanec}, {Pylypenko}, {Sitek}, {Howil}, {Zola}, {Michniewicz}, {Zejmo},
  {Lewis}, {Bronikowski}, {Potter}, {Andrzejewski}, {Merc}, {Street}, {Fukui},
  {Figuera Jaimes}, {Bozza}, {Rota}, {Cassan}, {Dominik}, {Tsapras},
  {Hundertmark}, {Wambsganss}, {B{\k{a}}kowska}, \& {S{\l}owikowska}}]{wu2024}
{Wu}, Z., {Dong}, S., {Yi}, T., {et~al.} 2024{\natexlab{b}}, \aj, 168, 62

\bibitem[{{Yang} {et~al.}(2024){Yang}, {Yee}, {Hwang}, {Qian}, {Bond}, {Gould},
  {Hu}, {Zhang}, {Mao}, {Zhu}, {Albrow}, {Chung}, {Kim}, {Park}, {Han}, {Jung},
  {Ryu}, {Shin}, {Shvartzvald}, {Cha}, {Kim}, {Kim}, {Lee}, {Lee}, {Lee},
  {Pogge}, {Zang}, {Abe}, {Barry}, {Bennett}, {Bhattacharya}, {Donachie},
  {Fujii}, {Fukui}, {Hirao}, {Itow}, {Kirikawa}, {Kondo}, {Koshimoto}, {Silva},
  {Li}, {Matsubara}, {Muraki}, {Suzuki}, {Tristram}, {Yonehara}, {Ranc},
  {Miyazaki}, {Olmschenk}, {Rattenbury}, {Satoh}, {Shoji}, {Sumi}, {Tanaka}, \&
  {Yamawaki}}]{yang2024}
{Yang}, H., {Yee}, J.~C., {Hwang}, K.-H., {et~al.} 2024, \mnras, 528, 11

\bibitem[{{Zang} {et~al.}(2020){Zang}, {Dong}, {Gould}, {Calchi Novati},
  {Chen}, {Yang}, {Li}, {Mao}, {Alton}, {Brimacombe}, {Carey}, {Christie},
  {Delplancke-Str{\"o}bele}, {Feliz}, {Gaudi}, {Green}, {Hu}, {Jayasinghe},
  {Koff}, {Kurtenkov}, {M{\'e}rand}, {Minev}, {Mutel}, {Natusch}, {Roth},
  {Shvartzvald}, {Sun}, {Vanmunster}, \& {Zhu}}]{zang2020}
{Zang}, W., {Dong}, S., {Gould}, A., {et~al.} 2020, \apj, 897, 180

\end{thebibliography}

\appendix

\section{Xallarap Models}
\label{sec:xallarap}

In this section, we explore whether the orbital motion of the source (xallarap effect) can explain the discrepancy between the position angle of the microlensed images measured using VLTI data and calculated from the light-curve model. Compared to the standard point-source point-lens model, models including the xallarap effect have seven additional parameters describing the shape of the source's orbit. These are $P_{\xi}$ -- the orbital period; $a_{\xi}$ -- the semimajor axis (expressed in Einstein radius units); $i_{\xi}$ -- the inclination of the orbit; $\Omega_{\xi}$ -- the longitude of the ascending node of the orbit; $u_{\xi}$ -- the argument of latitude at the reference epoch (which is taken to be $t_{0,\xi}=t_{0,\rm par}$); $e_{\xi}$ -- the eccentricity of the orbit; and $\omega_\xi$ -- the argument of periapsis of the orbit.

For simplicity, we consider only circular orbits, keeping $e_{\xi}=0$ and $\omega_\xi=0$. We also assume that the companion to the source is dark, given there is no strong evidence for the additional light in the VLTI data (Section~\ref{sec:luminous_blend_model}). We employ the \textsc{MulensModel} package by \citet{poleski2019} to calculate the source trajectory and magnification.

We search for best-fit xallarap models on a grid of orbital periods $P_\xi$ spanning from 10 to 251 days. We keep the orbital period fixed, but the remaining parameters ($a_{\xi}$, $i_\xi$, $\Omega_\xi$, $u_\xi$, and the standard microlensing parameters) are allowed to vary. The best-fit parameters are found using the MCMC approach \citep{foreman2013}. Simultaneously, we calculate the position angle of the minor image relative to the major image during the two epochs of VLTI observations (Section~\ref{sec:light_curve_model}). We assume uniform priors on all parameters.

The results of our calculations for the OGLE data are presented in Table~\ref{tab:xallarap}, where we report the $\chi^2$ statistic of the best-fit model and the predicted position angles. There is no strong evidence for xallarap in the data. The best-fit model, for the orbital period of $P_\xi =79.4$\,days, is favored over the model without xallarap by only $\Delta\chi^2=11.6$, which is not statistically significant given the increased number of free parameters. The predicted position angles are all similar to those calculated using the standard point-source point-lens model. However, their uncertainties are typically larger, which quantifies the additional degrees of freedom on the position of the source due to orbital motion.

\begin{deluxetable}{lrrr}[h]
\caption{Results of Xallarap Fits to the OGLE Data\label{tab:xallarap}}
\tablehead{\colhead{$P$ (days)} & \colhead{$\chi^2$} & \colhead{PA(1) (deg)} & \colhead{PA(2) (deg)}}
\startdata
No xallarap & 542.9 & $-168.37 \pm 1.83$ & $30.12 \pm 0.99$ \\
10.0 & 539.6 & $-168.75 \pm 1.57$  & $29.07 \pm 5.48$ \\
12.6 & 540.5 & $-168.69 \pm 1.30$  & $27.25 \pm 5.47$ \\
15.8 & 538.5 & $-168.73 \pm 1.63$  & $29.37 \pm 5.42$ \\
20.0 & 538.6 & $-168.51 \pm 1.37$  & $28.89 \pm 5.36$ \\
25.1 & 540.3 & $-168.11 \pm 1.45$  & $29.69 \pm 5.50$ \\
31.6 & 541.3 & $-168.24 \pm 1.42$  & $29.26 \pm 5.20$ \\
39.8 & 534.0 & $-168.79 \pm 2.46$  & $29.07 \pm 5.22$ \\
50.1 & 532.9 & $-169.33 \pm 2.87$  & $29.68 \pm 4.59$ \\
63.1 & 534.0 & $-169.34 \pm 3.86$  & $29.39 \pm 5.87$ \\
79.4 & 531.3 & $-169.42 \pm 2.80$  & $29.90 \pm 5.43$ \\
100  & 531.5 & $-169.85 \pm 10.16$ & $29.40 \pm 5.17$ \\
126  & 537.2 & $-169.39 \pm 29.20$ & $29.12 \pm 5.49$ \\
158  & 541.1 & $-168.72 \pm 6.08$  & $29.29 \pm 5.32$ \\
200  & 540.2 & $-168.40 \pm 29.12$ & $29.76 \pm 5.17$ \\
251  & 540.7 & $-168.44 \pm 3.96$  & $29.40 \pm 5.46$ \\
\enddata
\end{deluxetable}

\end{document}